\documentclass[
 reprint,
 superscriptaddress,
 amsmath,
 amssymb,
 aps,
 prd,
 floatfix,
 nofootinbib,
 fleqn
]{revtex4-1}

\usepackage[
colorlinks=true,
citecolor=blue,
filecolor=blue,
linkcolor=blue,
urlcolor=blue
]{hyperref}

\usepackage{graphicx}
\graphicspath{{./}{./figures/}}

\begin{document}

\title{A Method for Quantifying Position Reconstruction Uncertainty \\ in Astroparticle Physics using Bayesian Networks}

\author{Christina~Peters}
\email[]{petersc@udel.edu}
\affiliation{Department~of~Computer~and~Information~Sciences,~University~of~Delaware,~Newark,~Delaware,~USA.}

\author{Aaron~Higuera}
\affiliation{Department~of~Physics~and~Astronomy,~Rice~University,~Houston,~Texas,~USA.}

\author{Shixiao~Liang}
\affiliation{Department~of~Physics~and~Astronomy,~Rice~University,~Houston,~Texas,~USA.}

\author{Venkat~Roy}
\affiliation{Department~of~Electrical~and~Computer~Engineering,~Rutgers,~The~State~University~of~New~Jersey,~Piscataway,~New~Jersey,~USA.}

\author{Waheed~U.~Bajwa}
\affiliation{Department~of~Electrical~and~Computer~Engineering,~Rutgers,~The~State~University~of~New~Jersey,~Piscataway,~New~Jersey,~USA.}
\affiliation{Department~of~Statistics,~Rutgers,~The~State~University~of~New~Jersey,~Piscataway,~New~Jersey,~USA.}

\author{Hagit~Shatkay}
\altaffiliation{Deceased.}
\affiliation{Department~of~Computer~and~Information~Sciences,~University~of~Delaware,~Newark,~Delaware,~USA.}

\author{Christopher~D.~Tunnell}
\affiliation{Department~of~Physics~and~Astronomy,~Rice~University,~Houston,~Texas,~USA.}
\affiliation{Department~of~Computer~Science,~Rice~University,~Houston,~Texas,~USA.}

\collaboration{DIDACTS:~Data-Intensive~Discovery~Accelerated~by~Computational~Techniques~for~Science}

%\date{\today}

\begin{abstract}
\noindent Robust position reconstruction is paramount for enabling discoveries in astroparticle physics as backgrounds are significantly reduced by only considering interactions within the fiducial volume.
In this work, we present for the first time a method for position reconstruction using a Bayesian network which provides per interaction uncertainties.
We demonstrate the utility of this method with simulated data based on the XENONnT detector design, a dual-phase xenon time-projection chamber, as a proof-of-concept.
The network structure includes variables representing the 2D position of the interaction within the detector, the number of electrons entering the gaseous phase, and the hits measured by each sensor in the top array of the detector.
The precision of the position reconstruction (difference between the true and expectation value of position) is comparable to the state-of-the-art methods --- an RMS of 0.69~cm, $\sim$0.09 of the sensor spacing, for the inner part of the detector ($<$60~cm) and 0.98~cm, $\sim$0.12 of the sensor spacing, near the wall of the detector ($\geq$60~cm).
More importantly, the uncertainty of each interaction position was directly computed, which is not possible with other reconstruction methods. 
The method found a median 3-$\sigma$ confidence region of 11~cm$^2$ for the inner part of the detector and 21~cm$^2$ near the wall of the detector.
We found the Bayesian network framework to be well suited to the problem of position reconstruction. 
The performance of this proof-of-concept, even with several simplifying assumptions, shows that this is a promising method for providing per interaction uncertainty, which can be extended to energy reconstruction and signal classification.
\end{abstract}

\maketitle

\section{Introduction} \label{sec:intro}

A pioneering technology within astroparticle physics over the last decade has been noble-liquid time-projection chambers \cite{Aprile2010,DarkSide2018,LZ2020,PandaX2021,XENON2020,Aalbers2022}.
The detectors with this technology have been used in a range of searches for new physical phenomena ranging from dark matter to neutrinoless double-$\beta$ decay.
One of the reasons for their success in these fields is their excellent ability to discriminate background from signal of rare events, by reconstructing the type, energy and interaction position.  
For instance, the interaction position helps reduce the more than $10^8$ recorded events to less than $10^3$ events in the quiet central volume, often referred to as the fiducial volume \cite{XENON2018, XENON2019daq}.
State-of-the-art machine learning methods used to reconstruct the interaction position or energy of an interaction provide a single value, without per interaction uncertainty \cite{Lux2018, XENON2019analysis, Liang2022, PandaX2021, Simola2019}.

The Bayesian approach to probabilities can provide a framework for modeling and reasoning about uncertainty \cite{Bayes1763}.
Bayesian frameworks are used to handle uncertainty within data analysis and hypothesis testing.
The process of hypothesis testing often involves both a forward model and solving an inverse problem, which requires calculating causal factors from a set of observations or measurements \cite{James2006}.
Within particle physics, the forward model is often a simulation, whereas solving an inverse problem is referred to as reconstruction.
Complex simulations (e.g. G\textsc{eant}4 \cite{GEANT4} within high energy physics) that can act as forward models describe both the natural phenomena being measured and the experimental apparatus and are used to generate simulated data.
Detailed knowledge of the physical processes specific to a particle physics experiment are encoded in a forward model, leading to high-fidelity simulations (which can have millions or even billions of latent variables).
In these experiments, the likelihood for a given observation is often intractable. This means that the likelihood may be possible to know in theory, but in practice, it requires excessive computing resources to compute exactly, and thus requires approximation.

The process of statistical inference of an intractable likelihood is referred to as {\it likelihood-free inference} or {\it simulation-based inference} \cite{Cranmer2020}.
In this case where the likelihood is intractable, a typical approach is estimating the likelihood by calculating summary statistics and then comparing the experimental data to the simulated data (e.g.~using Approximate Bayesian Computation \cite{ABC1,ABC2}).
Neural networks, and machine learning algorithms in general, can increase the quality of the summary statistics used to estimate the likelihood, but generally do not estimate the joint probability distributions between the variables \cite{Radovic2018}.

There have been significant efforts applying and developing modern machine learning techniques within high energy physics \cite{Albertsson2018, hepmllivingreview}, as well as for neutrino experiments in particular \cite{Psihas2020}.
However, work on awareness of uncertainties or quantifying uncertainties when using machine learning techniques has been primarily within the context of jets or calibrating uncertainty estimates from deep learning \cite{Bollweg2019, Kasieczka2020, Estrade2019, Bellagente2021, Araz2021, Nachman2019, Koh2021}.

We chose the framework of Bayesian networks, which employs a graphical representation of probability distributions, to approach the problem of position reconstruction with uncertainties.
In this work, we demonstrate a method for position reconstruction, which provides not just uncertainties on the estimated position but a probability distribution over the possible positions, using the example of an astroparticle experiment aiming to directly detect dark matter.

\subsection{Probabilistic Graphical Models} \label{intro-PGMs}

A Probabilistic Graphical Model (PGM) can be used to represent the joint probability distribution over a set of variables using a graph as a data structure.
One of the first works to propose representing the interactions between variables with a graph structure was \cite{Gibbs1902}, where an undirected graph was used to represent the distribution over a system of particles.

Bayesian networks, one of the two broad classes of PGMs, use a \emph{directed acyclic graph} (DAG) to encode a probability distribution \cite{Howard1983, Smith1989} by making use of the independencies between the variables \cite{Verma1988, Geiger1998, Geiger1990a, Geiger1990b}.
The directed edges in Bayesian networks correspond to direct influence of one variable on another, which allows them to be used as interpretable models of physical systems for reasoning about causes and effects within the system \cite{Pearl1988, Wellman1990}.
Thus, this framework is well-suited for determining the posterior probability of any one of several possible known causes being a contributing factor to an observed event.
This framework also allows for a more compact representation of the probability distribution by utilizing the conditional independencies between the variables.
Applications of Bayesian networks are many and varied, with some of the earliest in genetics \cite{Elston1971} and medical diagnostic systems \cite{Heckerman1992}.

In Section~\ref{sec:methods}, we will introduce Bayesian networks and describe the simulated particle detector data used in this work. 
Next, in Section~\ref{sec:model}, the simulated data are used to construct a Bayesian network model of a particle detector with random variables representing the observable quantities of location, number of electrons, and sensor observables.
In Section~\ref{sec:positionreconstruction}, we perform inference on the Bayesian network to reconstruct interaction positions.
We then assess how well the model performs at correctly reconstructing particle interactions in Section~\ref{sec:results}, before concluding in Section~\ref{sec:conclusions}.

\section{Methods} \label{sec:methods}

\begin{figure}
\begin{center}
\includegraphics[width=0.46\textwidth]{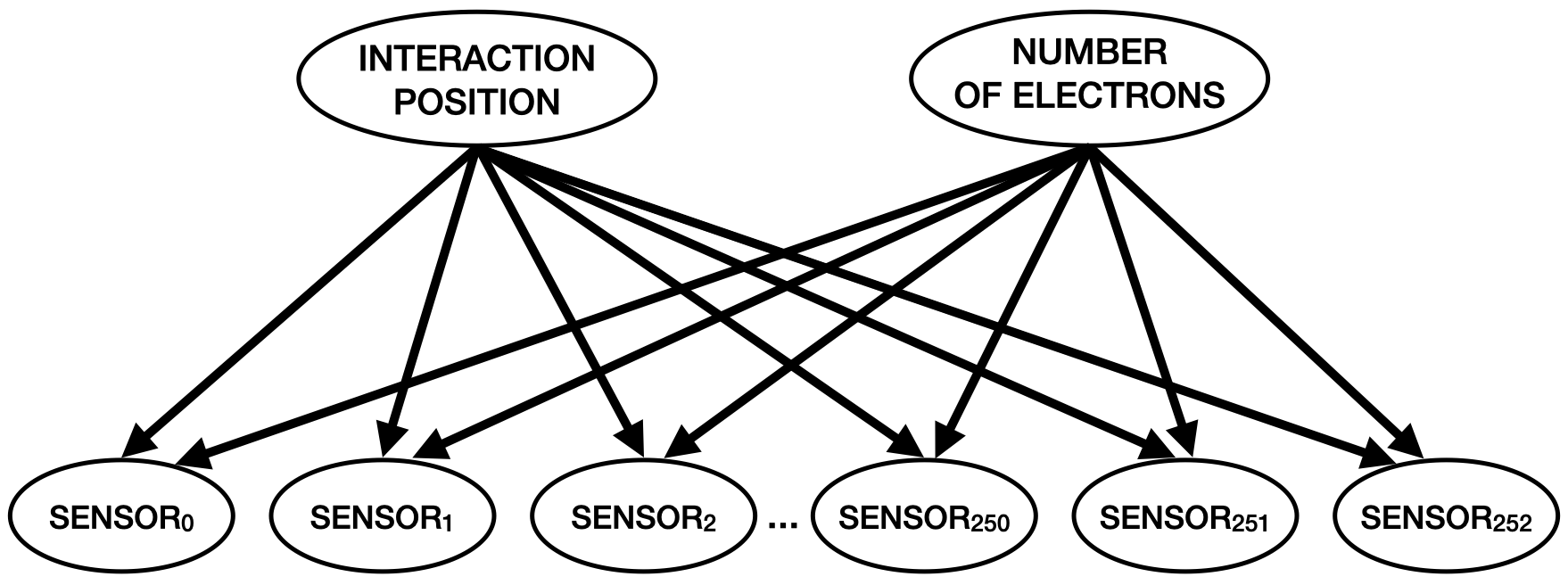}
\caption{Structure of the Bayesian network.
This graph structure provides a more compact representation of the high-dimensional joint probability distribution.
The structure implies the strong assumption that each {\it Sensor} node is conditional independent of all other {\it Sensor} nodes, given the {\it Interaction~Position} node and {\it Number~of~Electrons} node.
  \label{fig:model-structure}}
\end{center}
\end{figure}

A Bayesian network (shown in Figure~\ref{fig:model-structure}) is a graphical representation of random variables as nodes and arrows representing a causal relationship between variables.
As reconstruction is a challenge at the core of astroparticle physics analysis and the Bayesian network framework is a new approach, a detailed formalism is given here to aid others in utilizing this approach.
In addition, see \cite{Pearl1988, Jensen1996, Koller2009} for a thorough background on Bayesian networks.

{\em Random variables} represent a phenomenon which depends on a stochastic process, mapping from the sample space of all possible situations to the possible values the variables can take on.
A \emph{joint probability distribution} over two random variables denotes the probabilities of all combinations of the values of those variables. 
A probability model is completely determined by the \emph{full} joint probability distribution, the joint probability over all combinations of all the random variables in the model. 
This can be used to determine the probability of any outcome, but with more than a few random variables this is impractical to compute.
  
Let us consider a joint distribution over $n$ random variables $X_1,\ \dots\ , X_n$, where each variable has a set of $k$ possible discrete values.
One of the entries in the joint distribution would be $P(X_1=x_1,\ \dots\ , X_n=x_n)$, where there are $k^n$ such entries.
Unless this probability distribution is analytical, the size of the distribution can quickly become impractically large to perform computations or even store in memory.

A Bayesian network is a type of probabilistic graphical model that provides a means of overcoming this difficulty, assuming it is possible to represent the conditional independencies in the network structure.
In a Bayesian network each \emph{node} corresponds to a {\it random variable} and can be either discrete or continuous. 
\emph{Directed edges}, depicted as arrows connect pairs of nodes, where $X_1$ is a parent of $X_2$ if an arrow goes from $X_1$ to $X_2$. 
This graph must be acyclic (i.e., no cyclic paths) to avoid a variable depending upon itself. 
Additionally, each node has accompanying probability information to quantify the correlation of the parent nodes on the child node.

In many cases, the Bayesian network can concisely represent the full joint probability distribution by using the conditional independence properties of the distribution.
We can say that $X_1$ is conditionally independent of $X_2$ in the case that probability of $X_1$ given both $X_2$ and $X_3$, $P(X_1|X_2,X_3)$, is the same as the probability of $X_1$ given $X_3$, $P(X_1|X_3)$.
When $X_1$ is conditionally independent of $X_2$, learning the value of the variable $X_2$ contributes nothing to the certainty of the value of $X_1$.

The conditional independencies are encoded into the structure of the network (i.e., which nodes are connected by arrows).
The structure can be specified from domain knowledge or from observations of the system.
Once the structure is determined, then the local probability information for each node must be specified, again either based on expert knowledge of the domain or learned from data. 
This local probability information is a conditional distribution given the immediate parents of the node,
\begin{equation}\label{eq:BayesNet-localprobability}
{\bf P}(X_i | \mathrm{Parents}(X_i)).
\end{equation}
\noindent In the case of a discrete distribution without a parametric form, this can be a table where each row represents the conditional probability of each value of the node for each possible combination of values for the parent nodes. 
In the case of a continuous distribution, the distribution can be defined using a probability density function (e.g. Gaussian or Poisson).
In this case, it is necessary to store the parameters needed to specify the distribution (mean and standard deviation or expected value/rate).
For a discrete distribution the sum of conditional probabilities for all possible combinations of values for the parent nodes must be one.

In a Bayesian network, each entry in the joint distribution is defined as the product of the local conditional distributions,
\begin{equation}\label{eq:BayesNet-jointdistribution}
    P(X_1=x_1,\ \dots\ , X_n=x_n) = \prod\limits_{i=1}^{n} P(x_i | \mathrm{parents}(X_i)),
\end{equation} 
\noindent where $\mathrm{parents}(X_i)$ indicates the values of the parent nodes, $\mathrm{Parents}(X_i)$, that appear in $x_1,\ \dots\ , x_n$.

This graphical representation of the joint probability distribution over variables can be used to answer a query about the domain.
A \emph{probability query} is performed by computing the posterior probability distribution over the values of the {\it query variables}, ${\bf Y} = \{Y_{1},\ \dots,\ Y_{M}\}$, conditioned on the observed values, $\{x_{1},\ \dots,\ x_{N}\}$, of the {\it evidence  variables}, ${\bf X} = \{X_{1},\ \dots,\ X_{N}\}$,
\begin{equation}\label{eq:positionreconstruction-query}
  {\bf P}( Y_{1},\ \dots,\ Y_{M}\ |\ X_{1} = x_{1},\ \dots,\ X_{N} = x_{N} ).
\end{equation}
The purpose of the Bayesian network in this work is as a tool for position reconstruction, the process of inferring or estimating the location based on an observed hit pattern.
In Section~\ref{sec:model}, we will build a Bayesian network which can be used to perform position reconstruction by querying the joint distribution over all the random variables for the posterior probability distribution over interaction position. 

\subsection{Simulate Detector Data} \label{sec:Methods-simulations}

\begin{figure}
\begin{center}
\includegraphics[width=0.46\textwidth]{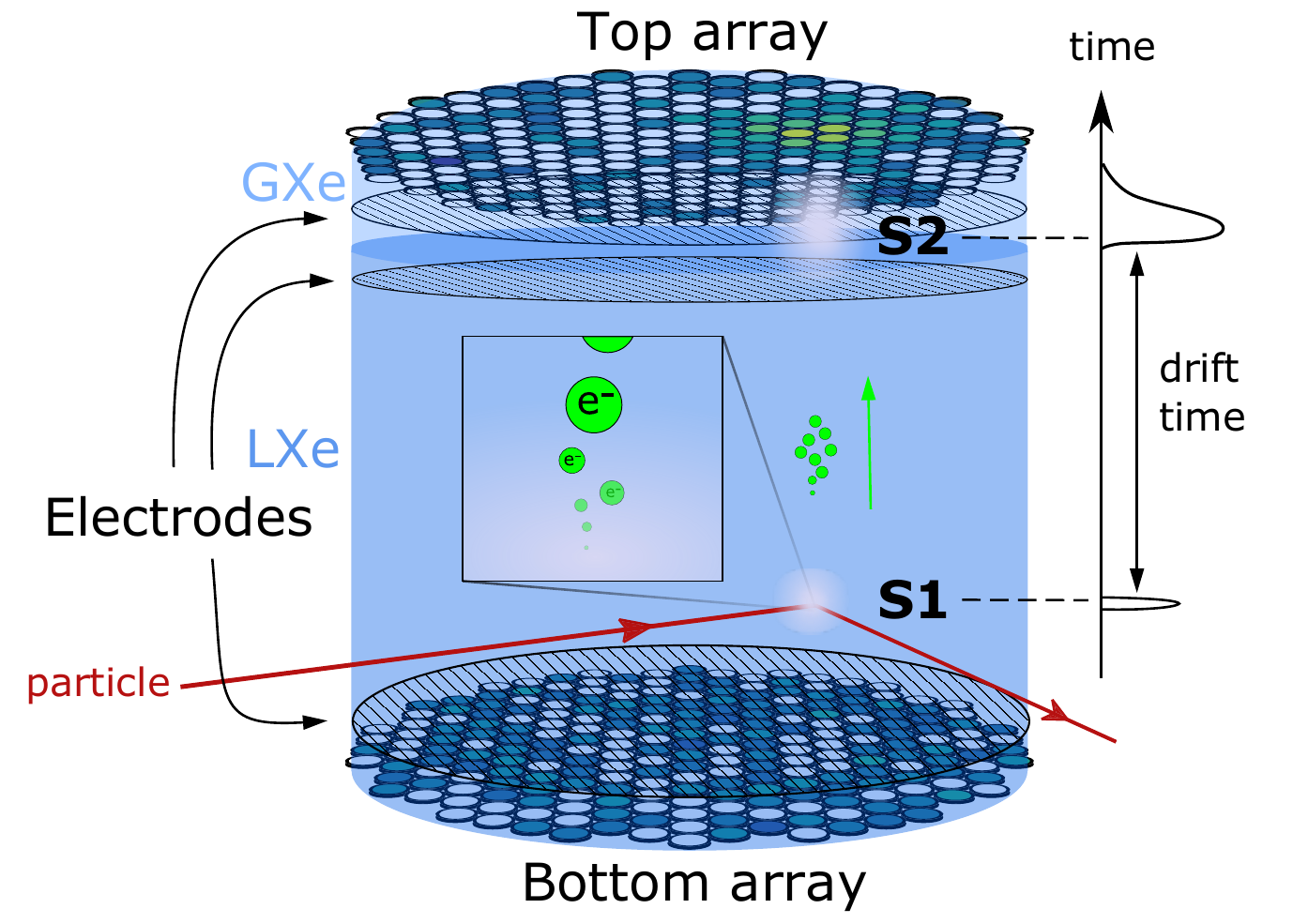}
\caption{A schematic of the working principle of a dual-phase liquid xenon TPC detector. Particles interact and deposit energy in liquid xenon target. An S1 signal is produced at the interaction location. Ionization electrons are released from the interaction point and drift toward the top of the TPC, where an S2 signal is produced and observed. From \cite{Liang2022}.
\label{fig:methods-xenonTPC}}
\end{center}
\end{figure}

\begin{figure*}
\begin{center}
\begin{tabular}{cc}
\includegraphics[width=0.4\textwidth]{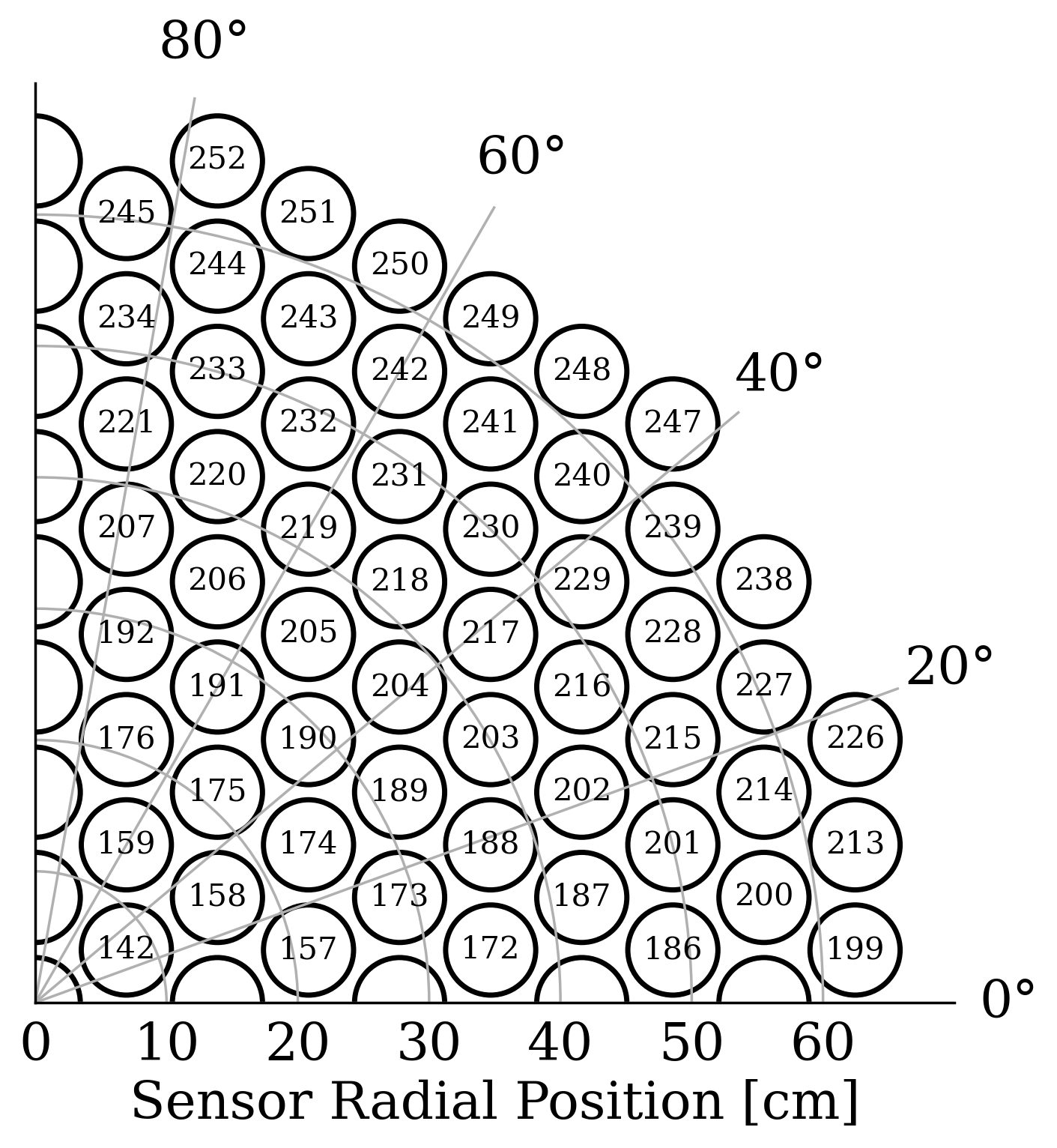} &
\includegraphics[width=0.4\textwidth]{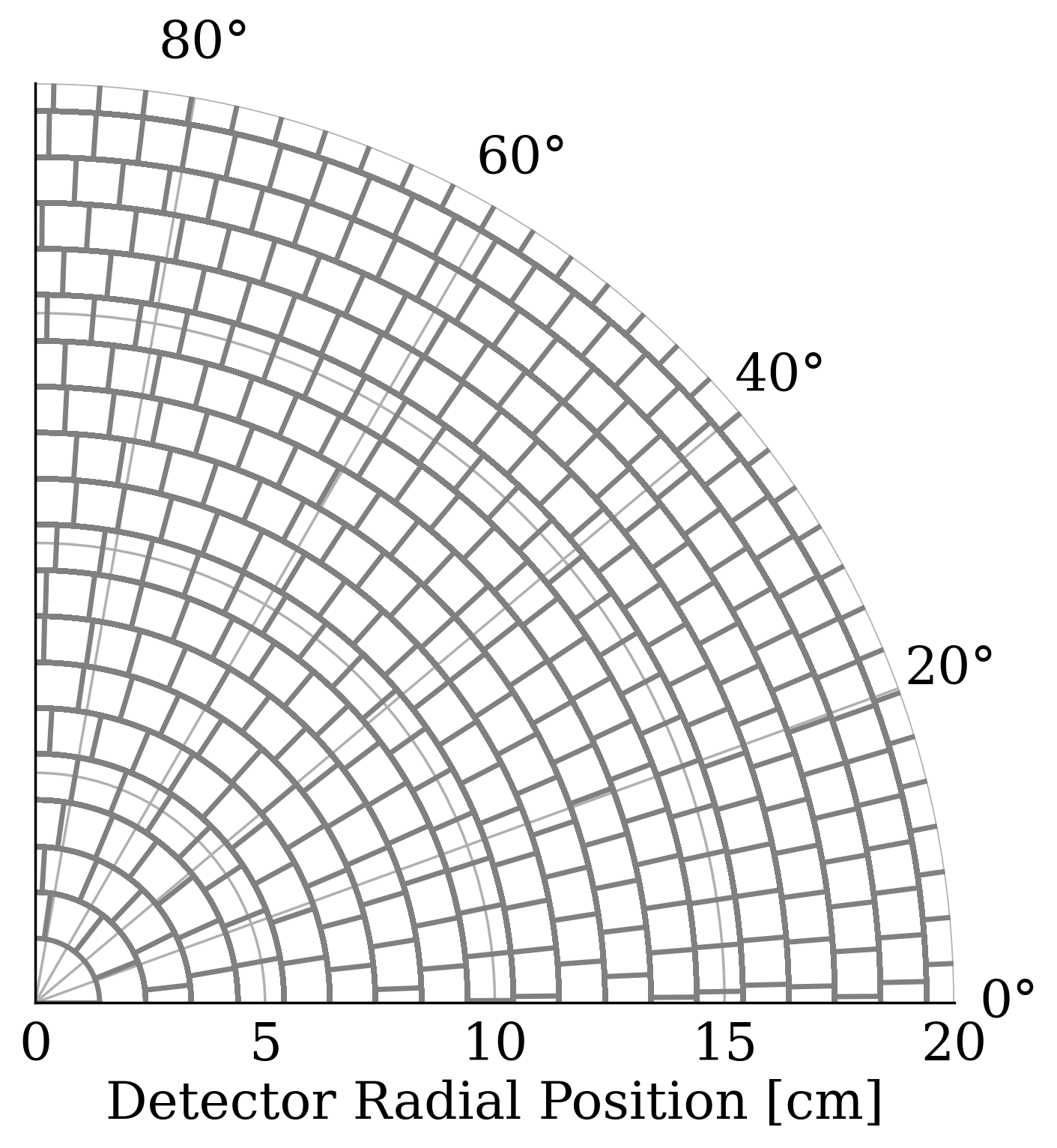}
\end{tabular}
\caption{Left, Positions of the sensors at the top of the detector. Only the first quadrant, containing 70 of the 253 total sensors, is shown in polar coordinates ($\rho$, $\phi$).
Right, boundaries of the discrete cells.
Only a portion of the first quadrant, containing 295 of the 13846 total cells, is shown. 
The discrete cells area was chosen to be 1~cm$^2$  as this is sufficiently small to provide precise position reconstruction.
In the Bayesian network, the cell within which an interaction occurs is described by the $C$ random variable, which has the set of values \{0, \dots , \ 13845\}.
\label{fig:methods-detectorcells}}
\end{center}
\end{figure*}

As a proof-of-concept, we will demonstrate the Bayesian network framework using the XENONnT detector design. 
See Figure~\ref{fig:methods-xenonTPC} for a schematic of the detector and sensor layout.
The XENONnT detector is a dual-phase xenon time-projection chamber (TPC) \cite{XENON2020}. The \emph{detector} contains 494 \emph{sensors}, photomultiplier tubes (PMTs), with 253 sensors on the top array and 241 on the bottom array arranged in the hexagonal structure shown in the left panel of  Figure~\ref{fig:methods-detectorcells}.
The spatial location of a particle interaction within the detector is denoted $\vec{l}$.
We approximate the detector as cylindrical in shape where, in cylindrical coordinates, $\vec{l} \in \{ 0\ \mathrm{cm} \leq \rho < 66.4\ \mathrm{cm},\ 0\ \mathrm{rad} \leq \phi < 2 \pi\ \mathrm{rad},\ -148.5\ \mathrm{cm}\ \leq z \leq 0\ \mathrm{cm}\}$.
The interaction positions will be estimated as a 2-dimensional position ($\rho, \phi$) using the sensors on the top array.
We leave the $z$ dimension for future work as it requires using the delay between the signal in the top array and bottom array.

In a dual-phase liquid xenon TPC, the 2-dimensional position is inferred from the observed products of the second scintillation in the gas phase.
When energetic particles scatter with the target atoms, the recoils excite and ionize the liquid xenon.
The xenon dimer produced then de-excites via the emission of scintillation light, which in the context of TPCs is referred to as the S1 signal.
The result of the ionization, free electrons, are drifted towards the top of the TPC where they are extracted into the gaseous phase by a strong electric field, causing another cascade that produces the second scintillation light, which is mostly measured by the top photosensor array given the geometrical arrangement.
This is referred to as the S2 signal.

The scintillation light observed by a sensor integrated over a short amount of time (from ten nanoseconds up to a microsecond) is referred to as a hit.
The collection of hits from all sensors is a \emph{hit pattern}.
The geometrical arrangement of the sensors instills position information within the hit patterns.
The hit pattern observed by the top array is denoted as $\vec{H}$, a 253 dimensional vector. 
The number of hits associated with the $j^{\mathrm{th}}$ sensor is a non-negative number, $H_{j}$.
This value is a float, rather than integer, because of sensor resolution.

Hit patterns are generated from the number of electrons, $e$, entering the gaseous phase.
The number of photons recorded by the sensors obey a Poisson distribution, where the expectation value is calculated by multiplying the light collection efficiency of each sensor and the number of photons in the S2 signal.

We generated a training set of $n= 5 \cdot 10^6$ S2 signals. 
The positions were generated from a uniform random distribution over the entire detector ($\{ 0\ \mathrm{cm} \leq \rho < 66.4\ \mathrm{cm},\ 0\ \mathrm{rad} \leq \phi < 2 \pi\ \mathrm{rad} \}$). 
The number of electrons extracted into the gas, $e$, were generated from a uniform random distribution from 1 to 2000, which includes the range of S2 signals cause by both dark matter particle interactions as well as common sources of background.
See Section~3 of \cite{Liang2022} for a detailed description of the data generation process.

In summary, the training data consists of hit patterns, each with its ground-truth position associated with a physical interaction and the ground-truth number of electrons, expressed as $\{\vec{H}^{i}, \vec{l}^{\,i}, e^{\,i} \}_{i}^{n}$.
One simulated hit pattern is shown in Figure~\ref{fig:methods-examplehitpatternposterior} (left) and additional examples are shown in Appendix~\ref{sec:appendix-examples}.

\begin{figure*}
\includegraphics[width=\textwidth]{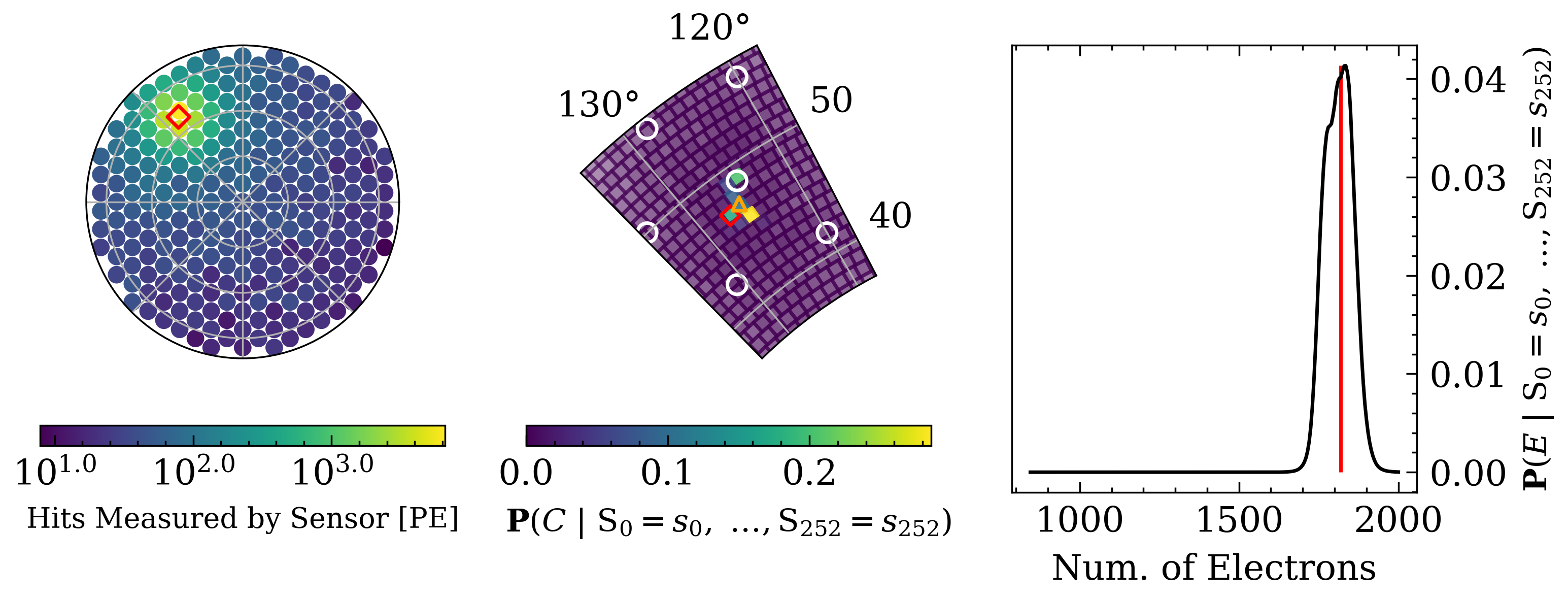}
\caption{Example of a simulated hit pattern (left), position reconstruction (center), and inferred number of electrons (right).
The ground-truth position of the example interaction, $\vec{l}$, simulated as described in Section~\ref{sec:Methods-simulations}, is shown as a red diamond in the left two panels.
On the left, the hit pattern shows the positions of 253 sensors in the top array of the detectors as filled circles, where the color of the filled circles indicates the intensity measured by the sensors.
In the center, the posterior probability distribution over the values of the $C$ random variable given the hit pattern at left, ${\bf P}(C\ |\ \mathrm{S_{0}}=s_{0},\ \dots ,\ \mathrm{S_{252}}=s_{252} )$, is visualized as a heatmap.
The expectation value of position is shown as an orange triangle and the sensor positions are shown as white circles.
The difference between the true position and the expectation value of position is 1.13~cm.
The 98.9108\% ($\sim$3-$\sigma$) and 99.9997\% ($\sim$5-$\sigma$) confidence intervals have areas of 17~cm$^2$ and 142~cm$^2$.
On the right, the posterior probability distribution over the values of the $E$ random variable given the hit pattern at left, ${\bf P}(E\ |\ \mathrm{S_{0}}=s_{0},\ \dots ,\ \mathrm{S_{252}}=s_{252} )$, is shown.
The ground-truth number of electrons of the example interaction, $e$ = 1819, is shown as a vertical red line.
\label{fig:methods-examplehitpatternposterior}}
\end{figure*}

\section{Represent a Particle Detector as a Bayesian Network} \label{sec:model}

The goal in building this Bayesian network is to learn the underlying distribution from which the data, $\{\vec{H}^{i}, \vec{l}^{\,i}, e^{\,i} \}_{i}^{n}$, is sampled.
The construction of the Bayesian network begins with defining the random variables.
The random variables describe the location of the particle interactions, the number of electrons generated at the second scintillation (S2), as well as the hits measured by each sensor in the hit pattern.
The random variables will be represented as a collection of nodes in the Bayesian network.
We represent the two dimensional position of the interaction with the {\it Interaction~Position} random variable which maps the $i$-th particle interaction to the position of its ground-truth spatial location, $\vec{l}^{i}$.
We represent the number of electrons produced by the interaction with the {\it Number~of~Electrons} random variable which maps the $i$-th particle interaction to the ground-truth number of electrons, $e^{\,i}$.
In addition, we define a set of 253 random variables,
\begin{equation}
\textbf{\textit{Sensor}} = \{\textit{Sensor}_{0}, \dots, \textit{Sensor}_{252}\},
\end{equation}
\noindent which maps the $i$-th particle interaction to the ground-truth hits measured by each sensor, $H^{i}_{j}$.

Now that we have defined the nodes, we will discuss the edges of the graph.
The edges connecting pairs of nodes correspond to the direct dependencies between the random variables.
An edge between two nodes can be represented with a \emph{conditional probability distribution} (CPD) of the parent node given the child node, which will be learned from the data.
The presence (and absence) of edges between nodes can be determined from knowledge about the system being represented or learned from data.
We use the graph structure shown in Figure~\ref{fig:model-structure} as a proof-of-concept.
While this is a simple graph structure, it has been proven to be quite effective in practice.

The {\it Interaction~Position} node and {\it Number~of~Electrons} node are the two parents of all of the {\it Sensor} nodes. 
From the {\it Interaction~Position} node there is an arrow to each {\it Sensor} node, indicating a direct dependence of the sensor values on the location of the particle interaction.
Similarly, there is an arrow from the {\it Number~of~Electrons} node to each {\it Sensor} node.
Therefore, every sensor contains information on position and the number of electrons produced in the interaction.

This structure assumes that the {\it Sensor} nodes are conditionally independent of each other, given the {\it Interaction~Position} and {\it Number~of~Electrons} nodes. 
That is, for an interaction at a given position in the detector with a given number of electrons, the probability distributions of the sensors are independent of each other.
The sensor random variables are not strictly independent given the {\it Interaction~Position} and {\it Number~of~Electrons} random variable, but without this simplifying assumption it would be necessary to consider dependencies between all unique pairs of sensors.
In practice, this graph structure has been shown be effective for classification, even in cases where independence assumptions are violated, with the advantages of being easier to interpret, faster to learn, faster to query, and smaller to store in memory \cite{Domingos1997}.

\subsection{Data Representation} \label{sec:model-datarepresentation}

We will now modify the representation of the random variables by either changing the continuous variables to discrete, or parameterizing the continuous variables with a function.
The advantage to discrectizing a continuous random variable is that the conditional probabilities can be specified explicitly for each value, without defining a parametric form.

Locations within the detector can be represented as discrete values by pixelizing the area into cells. 
We divide the detector into discrete cells which each have an area of 1 $\mathrm{cm}^{2}$ (approximately square in shape and $\sim$1~cm on a side), with the exception of the central cell, which has an area of $\sim$6.2~cm$^{2}$.
The size of the cells was chosen so the position reconstruction will have the necessary precision to be scientifically useful.
The discrete cells with boundaries in radius and angle ($\rho$, $\phi$) are shown in the right panel of Figure~\ref{fig:methods-detectorcells}.

Rather than the {\it Interaction~Position} random variable, we represent the locations with a single discrete multinomial random variable, $C$, which maps the $i$-th particle interaction to the cell, $c$, of its ground-truth spatial location, $\vec{l}^{\,i}$. 
The set of possible values that the random variable $C$ can take is $\{ 0,\ \dots,\ 13845\}$, which are indices corresponding to the discrete cells. 
Assuming the particle interactions occur uniformly within the detector, the prior probability of an interaction occurring within a cell is the fraction of the area of the detector within that cell.

Instead of the {\it Number~of~Electrons} random variable, we represent the number of electrons produced by the interaction with a single random variable, $E$, which maps the $i$-th particle interaction to the ground-truth number of electrons, $e^{\,i}$.
The set of possible values that it can take is all positive integers.

The number of hits measured by a sensor, which was previously represented by the set of random variables \textbf{\textit{Sensor}}, is now represented as a continuous function by fitting the training data with a suitable distribution and is now represented by the set of random variables ${\bf S}  = \{\textit{S}_{0}, \dots, \textit{S}_{252}\}$.
Each random variable $S_j$ will be represented by a \emph{Poisson distribution} which is commonly used to represent count data, such as the number of photons arriving at the sensor over a given time period.
This is a distribution over non-negative integers and has one parameter $\lambda$, which is both the expected value and the variance.
A random variable $X$ sampled from Poisson($\lambda$) has the distribution
\begin{equation}\label{eq:detector-Poisson}
P(X=x) = \frac{\exp{(-{\lambda})}\ \ {\lambda}^{x}}{x!}.
\end{equation}
As the set of possible values that this distribution can take is all non-negative integers, the number of hits are rounded to the nearest integer value.

\subsection{Learn the Conditional Probabilities} \label{sec:model-conditionalprobabilities}

Now that the structure of the graph and the random variable have been defined, the next step is learn the parameters of the network from the data. 
Specifically, we determine the conditional probability distributions (CPDs) for each random variable.

The {\it C} node ({\it Interaction~Position} in Figure~\ref{fig:methods-detectorcells}) has no parent nodes and thus the probability distribution is a prior --- not dependent on other random variables.
We define this to be the fraction of interactions in the training set within each discrete cell,
\begin{equation}\label{eq:learningmodel-prior}
    \mathrm{P}(C\ =\ c)\ =\ \frac{\mathrm{Interactions\ in\ Cell\ } c}{\mathrm{Total\ Interactions}}.
\end{equation}
\noindent We define the prior distribution over the $E$ node ({\it Number~of~Electrons} in Figure~\ref{fig:methods-detectorcells}) to be uniform.

In this structure all the ${\bf S}$ nodes ({\it Sensor} in Figure~\ref{fig:methods-detectorcells}) have both the $C$ node and the $E$ node as parents.
Therefore, the CPD of each $S$ node depends on the $C$ node and the $E$ node.
For the $j$-th sensor, the CPD is ${\bf P}(S_{j}\ |\ C,\ E)$. 

\begin{figure}
\begin{center}
\includegraphics[width=0.45\textwidth]{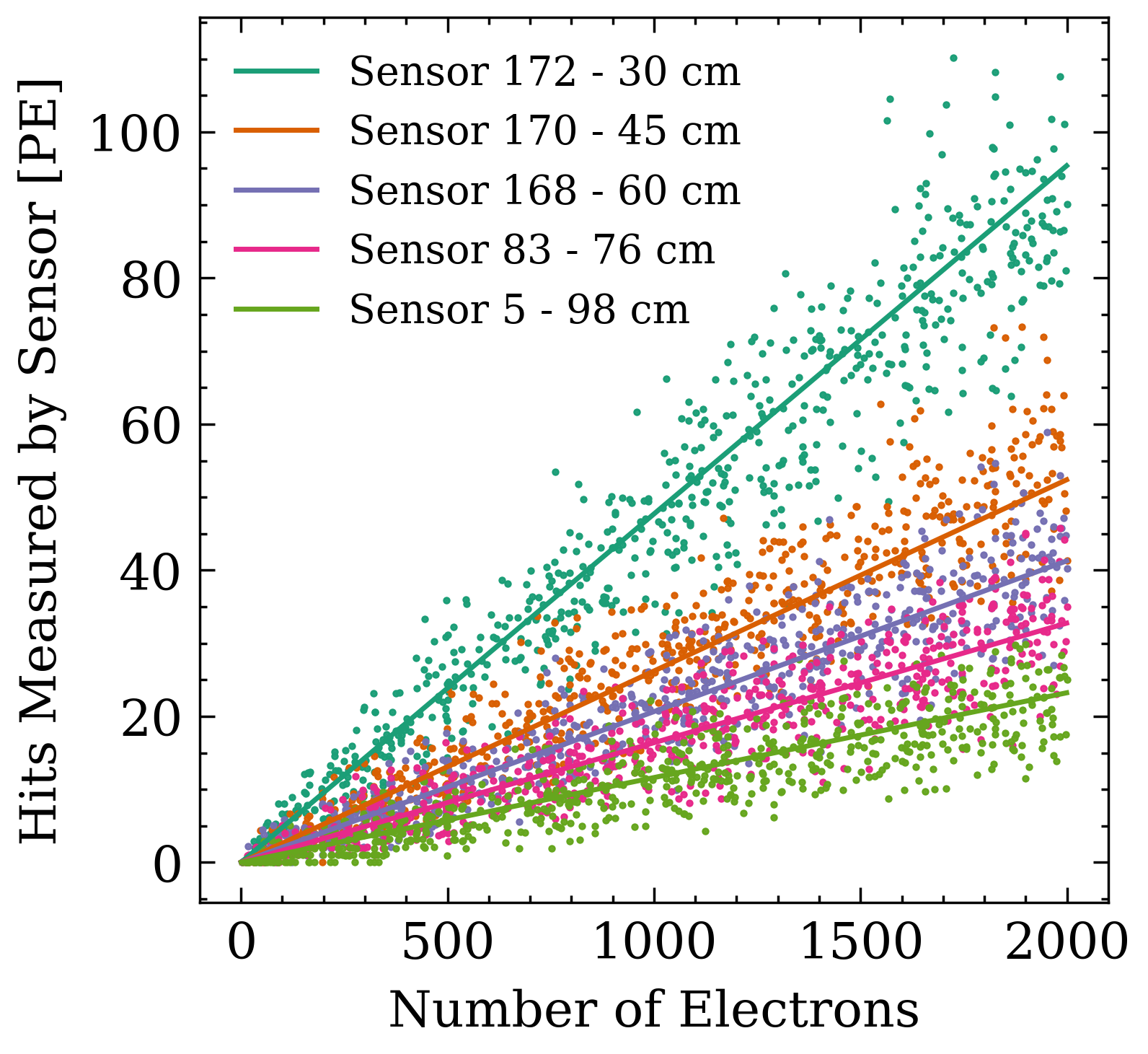}
\caption{Dependence between the set of  nodes ${\bf S}$ and the $E$ node. 
All interactions were generated within the cell $C$ = 10168 (centered at $\rho, \phi$ = 33.90~cm, 58.54$^\circ$). 
The color of the scatter points indicates hits measured by different sensors where the distance between the location of the interactions and the sensors is noted in the legend.
The lines are linear fits to the scatter points with the $y$-intercept fixed at 0.
\label{fig:training-data-fits}}
\end{center}
\end{figure}

In Figure~\ref{fig:training-data-fits}, the scatter points show a set of simulated interactions which were generated within the cell $C$ = 10168 (centered at $\rho, \phi$ = 33.90~cm, 58.54$^\circ$) with the number of electrons drawn from a uniform distribution from 1 to 2000.
The slope, $m$, was determined by fitting a straight line using non-linear least squares with the $y$-intercept fixed at 0.
A $m_{c,j}$ was defined for each sensor node $S_j$ in ${\bf S}$ and all possible values $c$ of the $C$ node.

\begin{figure}
\begin{center}
\includegraphics[width=0.45\textwidth]{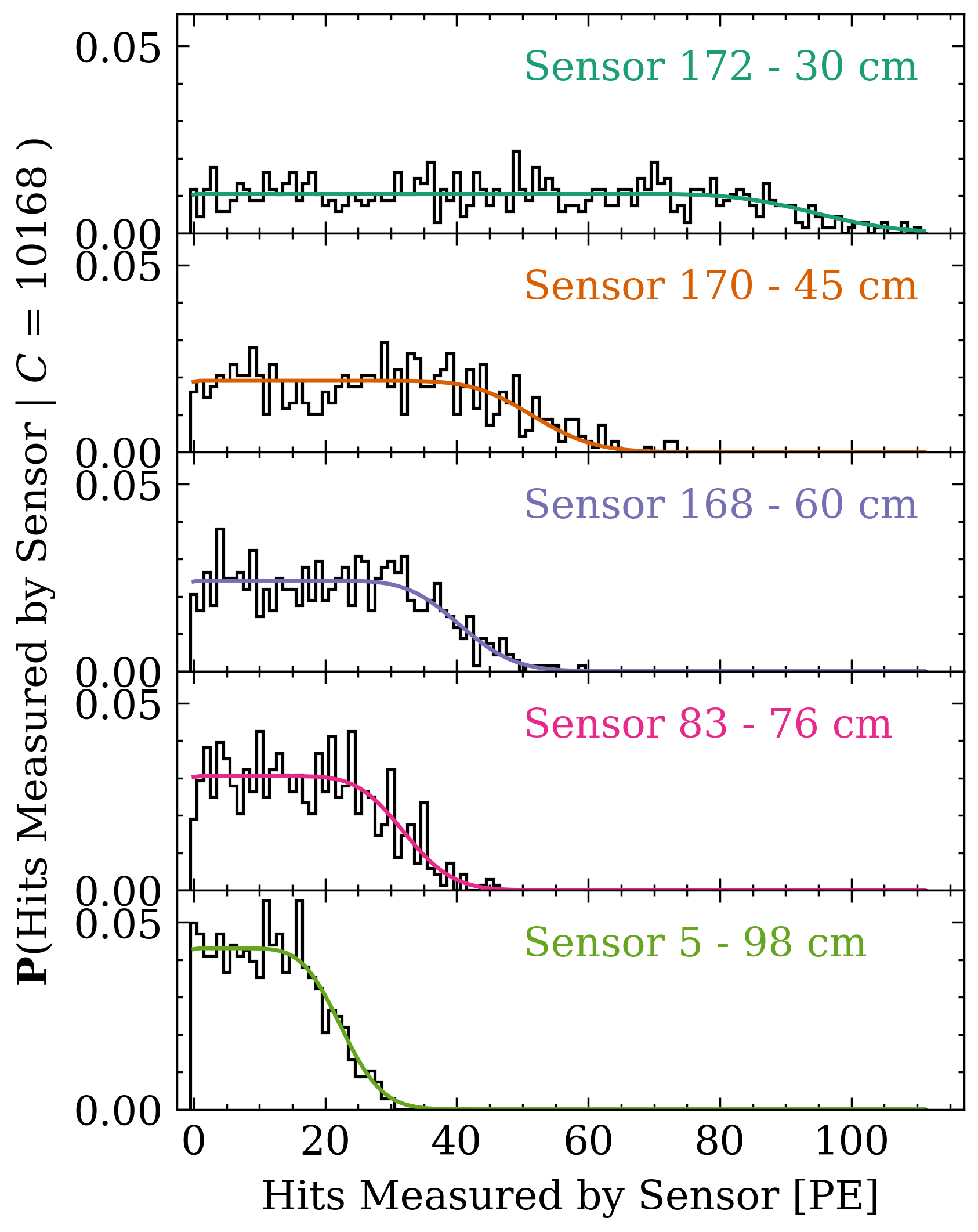}
\caption{Parameterizing the set of random variables ${\bf S}$ as Poisson distributions. 
The black normalized histograms in each panel show the number hits measured by sensors for interactions generated within cell $C$ = 10168 (centered at $\rho, \phi$ = 33.90~cm, 58.54$^\circ$).
The colored lines show the conditional probability distribution, after summing over all values of the $E$ node. 
Thus, this is conditional probability that the node $S_j$ has the intensity $s_j$ given $C$ = $c$, as defined in Eq.~\ref{eq:learningmodel-conditionalpoisson1}.
\label{fig:training-data-fits-whist}}
\end{center}
\end{figure}

We define the CPD, given any assignment to the $C$ node and $E$ node, to be
\begin{equation}\label{eq:learningmodel-conditionalpoisson1}
    P(S_j = s_j\ |\ C = c,\ E = e) = \frac{\exp{(-{\lambda_{c,j,e}})}\ \  {(\lambda_{c,j,e})^{s_j}}}{s_j!},
\end{equation}
\noindent where
\begin{equation}\label{eq:learningmodel-conditionalpoisson2}
    \lambda_{c,j,e} = e \cdot m_{c,j}.
\end{equation}
\noindent Each panel of Fig~\ref{fig:training-data-fits-whist} shows $P(S_j = s_j\ |\ C = c,\ E = e)$ for interactions within the cell {\it C} = 10168, after summing over all values of the {\it E} node, for a single sensor.

The total number of values stored to calculate the CPDs using this graph structure and parameterization of the random variables is the number of discrete cells $\times$ number of sensors (13846 $\times$ 253).

\section{Position Reconstruction} \label{sec:positionreconstruction}
 
The purpose of the Bayesian network defined in the previous section is as a tool for position reconstruction, the process of inferring or estimating the location based on an observed hit pattern.
As defined in Equation~\ref{eq:positionreconstruction-query}, a Bayesian network can be queried by computing the posterior probability distribution over the values of the query variables conditioned on the observed values of the evidence  variables.
This Bayesian network can be used to perform position reconstruction by querying the posterior probability distribution over the values of the random variable that represents the interaction position conditioned on the observed values of the  variables which represent the sensors.
Therefore, the query variable is the $C$ node and the evidence variables are the ${\bf S}$ nodes. 
The posterior probability distribution over $C$, given the observed values of the ${\bf S}$ nodes, $s_{0},\ \dots \ ,\ s_{252}$, is
\begin{equation}\label{eq:positionreconstruction}
  {\bf P}(C |\  \mathrm{S_{0}} =  s_{0},\ \dots,\ \mathrm{S_{252}}=s_{252}).
\end{equation}
\noindent The ${\bf P}$ of Equation~\ref{eq:positionreconstruction} is a categorical probability distribution which assigns the probability for the values of the $C$ random variable. 
As there are observed values for all of the ${\bf S}$ nodes, the only non-query and non-evidence variable to be marginalized over in the probability query is the $E$ node.

For the simulated data described in Section~\ref{sec:Methods-simulations}, the observed value of $j$-th sensor, $s_j$, associated with the $i$-th particle interaction is $\vec{H}_{j}^{i}$.
The posterior for the $i$-th particle interaction in the test set, where the observed value of $j$-th sensor is $s_j$, can be written as
\begin{multline}\label{eq:positionreconstruction-posterior}
  {\bf P}^{i} (C\ |\ \mathrm{S_{0}}=\vec{H}_{0}^{i},\ \dots,\ \mathrm{S_{252}}=\vec{H}_{252}^{i} ) = \\
  \sum\limits_e {\bf P}(C,\ E = e) \prod\limits_j\ {\bf P}(S_j = \vec{H}_{j}^{i},\ C,\ E = e),
\end{multline}
\noindent where 
\begin{equation}\label{eq:positionreconstruction-posterior-normalization}
  \sum\limits_c \ {\bf P}^{i} (C = c \ |\ \mathrm{S_{0}}=\vec{H}_{0}^{i},\ \dots ) = 1.
\end{equation}
\noindent The posterior over $E$ can be calculated in a similar manner by marginalizing over $C$, as well as the joint probability distribution over $C$ and $E$.

We used a test set of 50,000 particle interactions which were simulated using the same parameters as the training data.
Examples of the posterior over $C$ and posterior over $E$ are shown in Figure~\ref{fig:methods-examplehitpatternposterior} and additional examples are shown in Appendix~\ref{sec:appendix-examples}.

\section{Model Performance} \label{sec:results}

This network is an approximation of the joint probability distribution over hit patterns, interaction locations, and number of electrons.
We write set of a metrics to measure how well the network performs position reconstruction.
These metrics will measure the difference between the ground-truth spatial locations, $\vec{l}$, and the posterior probability distribution over $C$ given the hit pattern (the observed values of the ${\bf S}$ nodes) for a test set of labeled simulated data, ${\bf P}(C\ |\ \mathrm{S_{0}}=s_{0},\ \dots ,\ \mathrm{S_{252}}=s_{252} )$.
 
\subsection{Coverage Probability} \label{sec:results-coverageprobability}

The purpose of this metric is to determine if the posterior probability distributions calculated in the previous section can provide confidence intervals with good performance (i.e., the coverage of the confidence interval is close to the intended confidence level).
This will assess, for example, if the ground-truth locations fall within the region of the detector with a posterior probability of 0.95 for 95\% of the interactions in the test set.

However, because the position is divided into discrete cells it is not possible to define regions of the detector with a 95\% confidence interval for each interaction in the test set.
To address this, the metric considers a cumulative probability distribution over the cells ordered from highest to lowest probability.
The $j$-th highest probability cell in the probability distribution of the $i$-th interaction in the test set is defined as ${c}^{i}_{j}$ and the cell with the highest probability for the interaction to have occurred is denoted ${c}^{i}_{1}$. 
The coverage of the confidence interval for the $i$-th interaction in the test set is the sum of the probabilities of the cells with the $j$-th highest probabilities,
\begin{equation}\label{eq:metrics-averageP}
P^{i}_{j}  = \sum\limits_{j = 1}^{j}\ P^{i} ({\it C}\ = {c}^{i}_{j}|\ \mathrm{S_{0}}=s_{0},\ \dots) .
\end{equation}

In the example position reconstruction shown in Figure~\ref{fig:methods-examplehitpatternposterior}, the cell with the highest probability for the interaction to have occurred is $P_{1} = 0.28$, therefore the 28\% confidence interval has an area of 1~cm$^2$.
The sum of the probabilities of the cells with the 17 highest probabilities is $P_{17} = 0.9891$, therefore the 98.91\% confidence interval has an area of 17~cm$^2$.

\begin{figure}
\begin{center}
\includegraphics[width=0.46\textwidth]{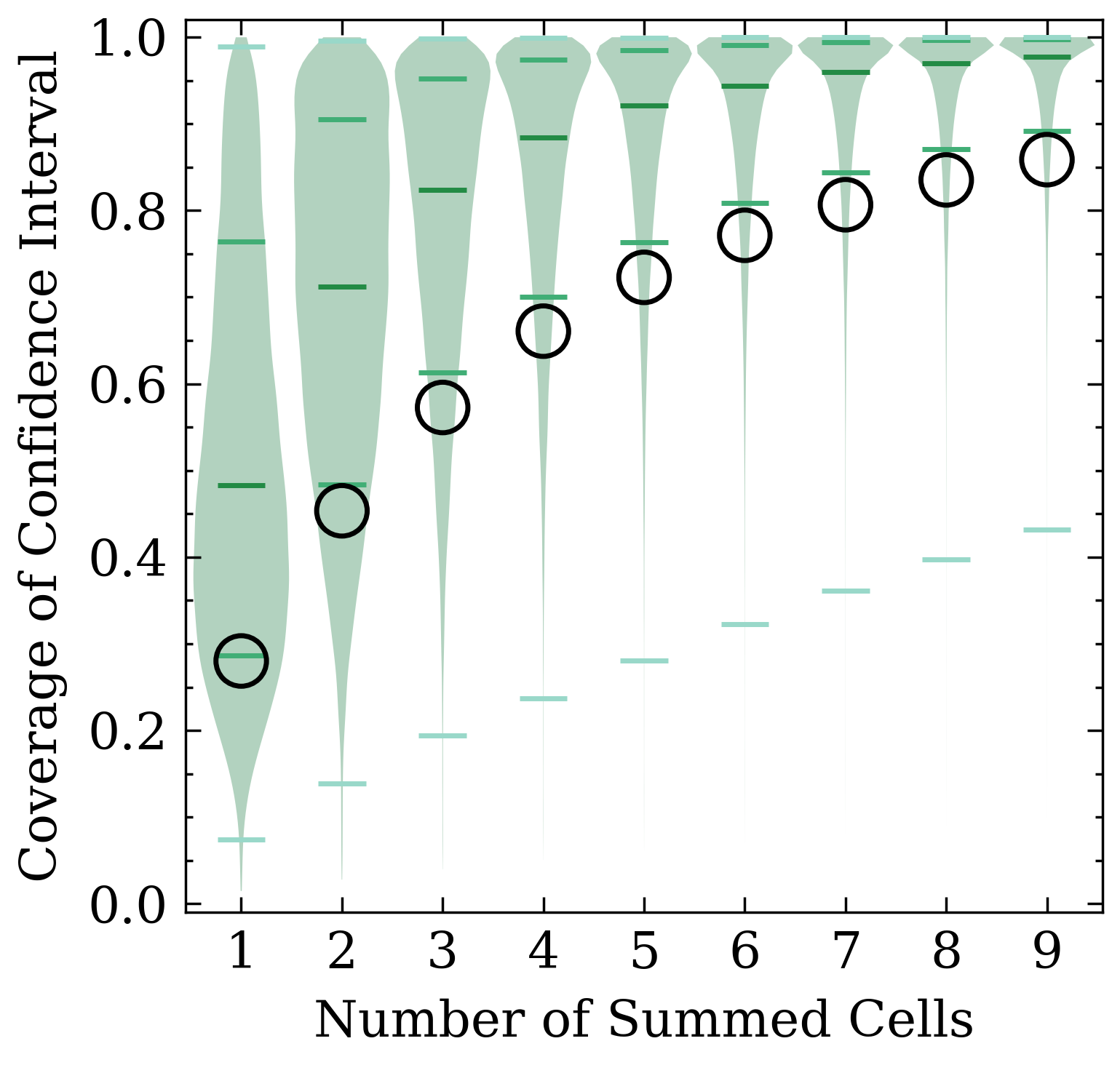}
\caption{
The x-axis indicates the number of summed cells where the highest probability cell (${c}^{i}_{1}$) is at farthest left and the sum over the 9 highest probability cells is shown at farthest right. 
The black circle is the fraction of the interactions where the summed cells contain the ground-truth position, as defined in Equation~\ref{eq:metrics-proportion}. 
The green violin shows the coverage of the confidence interval as defined in Equation~\ref{eq:metrics-averageP} for all interactions in the test set.
The green whiskers indicate the median, as well as the 1-$\sigma$ and 3-$\sigma$ spread for the coverage of the confidence interval.
This figure demonstrates that the actual coverage of the confidence interval is greater than the intended coverage and therefore the confidence intervals generated by this model are anti-conservative.
\label{fig:results-coverageprobability}}
\end{center}
\end{figure}

The cell that contains the ground-truth spatial location, $\vec{l}^{\ i}$, is defined as $\dot{c}^{i}$.
We define the fraction of the interactions where the $j$-th highest probability cell contains $\vec{l}^{\ i}$ in the test set of $n_{\mathrm{test}}$ interactions as
\begin{equation}\label{eq:metrics-proportion}
F_{j} = \sum\limits^{n_{\mathrm{test}}}_{i=0} \bigg\{ \genfrac{}{}{0pt}{}{\frac{1}{n_{\mathrm{test}}}\ \mathrm{if\ }\dot{c}^{i}\ \in \{{c}^{i}_{1},\ \dots,\ {c}^{i}_{j} \}}{0\ \mathrm{otherwise}} .
\end{equation}
\noindent In the example shown in Figure~\ref{fig:methods-examplehitpatternposterior}, the cell that contains the $\vec{l}^{\ i}$ is the cell with the second highest probability, ${c}_{2}$.
The ground-truth spatial location occurred within the cells with the 2 highest probabilities for approximately 46\% of the interactions in the test set.

On the far left of Figure~\ref{fig:results-coverageprobability}, we compare the coverage of the confidence interval of the cell with the highest probability for all interactions in the test set (shown as a green violin plot) and the fraction of the interactions where the highest probability cell contains the ground-truth location (shown as a black circle).
This shows that the actual coverage of the confidence interval is higher than the intended coverage --- half of the interactions in the test set have $P_1 > 0.475$, but for only 28\% of interactions was the ground-truth spatial location within ${c}_{1}$.
Figure~\ref{fig:results-coverageprobability} shows that this trend is consistent for confidence intervals up to 9 summed cells and that the actual coverage of the confidence interval is greater than the intended coverage for approximately 84\% of the interactions in the test set.

The confidence intervals generated by this model can be described as over-confident; this is termed an anti-conservative confidence interval.
Both the conditional independence assumptions in the Bayesian network structure and the use of a discrete distribution to represent interaction position can produce a discrepancy between the actual coverage of the confidence interval and the intended coverage.
In practice, a scaling factor could be used to partially resolve this undercoverage.
Additionally, a more complex graph structure that takes in to consideration dependencies between sensor nodes may provide confidence intervals with better performance.

\subsection{Difference between Ground-Truth and Inferred Position} \label{sec:results-expectationvalue}

The purpose of this metric is to calculate the spatial distance between the ground-truth location and the inferred location for a test set of interactions. 
This is primarily to compare to existing results from current position reconstruction methods.

We start by calculating the expectation value of the position from the posterior probability distribution over the values of the $C$ node.

The expectation value of a random variable $X$ with a set of $n$ possible values $\{x_1,\ \dots,\ x_n\}$ is defined by 
\begin{equation}
\langle X \rangle = \sum\limits_{j=1}^{n} {\bf P}_j x_j ,
\end{equation}
\noindent where ${\bf P}$ is the probability mass function and 
\begin{equation}
\sum\limits_{j=1}^{n} {\bf P}_j = 1.
\end{equation}
\noindent The center of each cell is specified by positions along the radial and angular axes, $\rho$ and $\phi$.
For the radial direction, $\rho$, the standard definition of expectation value above is sufficient. 

However, for cyclic quantities, such as the random variable $\phi$ which is an angle, the usual calculation of a mean is not appropriate.
A circular mean is a mean designed for cyclic quantities such as angles and hours of the day.
For example, the arithmetic mean of $0^\circ$ and $360^\circ$ is $180^\circ$, which is inconsistent with the circular mean of $0^\circ$ or  $360^\circ$.

One approach to calculating a circular mean or circular expectation value is to use complex numbers. 
We define the angular position of an interaction represented as a complex number to be $z = \exp{(i \cdot \phi)}$, where $i$ is the unit imaginary number.
The expectation value of $\phi$, $\langle \phi \rangle$, is the argument of the resultant vector,
\begin{equation}\label{eq:metrics-expectationvaluecircularposition}
\langle \phi \rangle  = \mathrm{Arg} \Big(\langle z \rangle \Big)= \tan^{-1} \Bigg( \frac{\mathrm{Im}  \Big(\langle z \rangle  \Big)}{\mathrm{Re}  \Big( \langle z \rangle  \Big)} \Bigg) ,
\end{equation}
\noindent where $\langle z \rangle$ is the expectation value of $z$. The difference between the ground-truth location, ${\vec l}^i = \{\rho^i, \phi^i \}$, and the expectation value of the position, ${\langle \vec l \rangle}^i = \{{\langle \rho \rangle}^i, {\langle \phi \rangle}^i \}$, in Cartesian coordinates is
\begin{equation}
\begin{split}
{\Delta x}^i = \rho^i \sin{\big( \phi^i \big)} - {\langle \rho \rangle}^i \sin{\big({\langle \phi \rangle}^i \big)}, \\
{\Delta y}^i = \rho^i \cos{\big( \phi^i \big)} - {\langle \rho \rangle}^i \cos{\big({\langle \phi \rangle}^i \big)}.
\end{split}
\end{equation}

\begin{figure}
\includegraphics[width=0.46\textwidth]{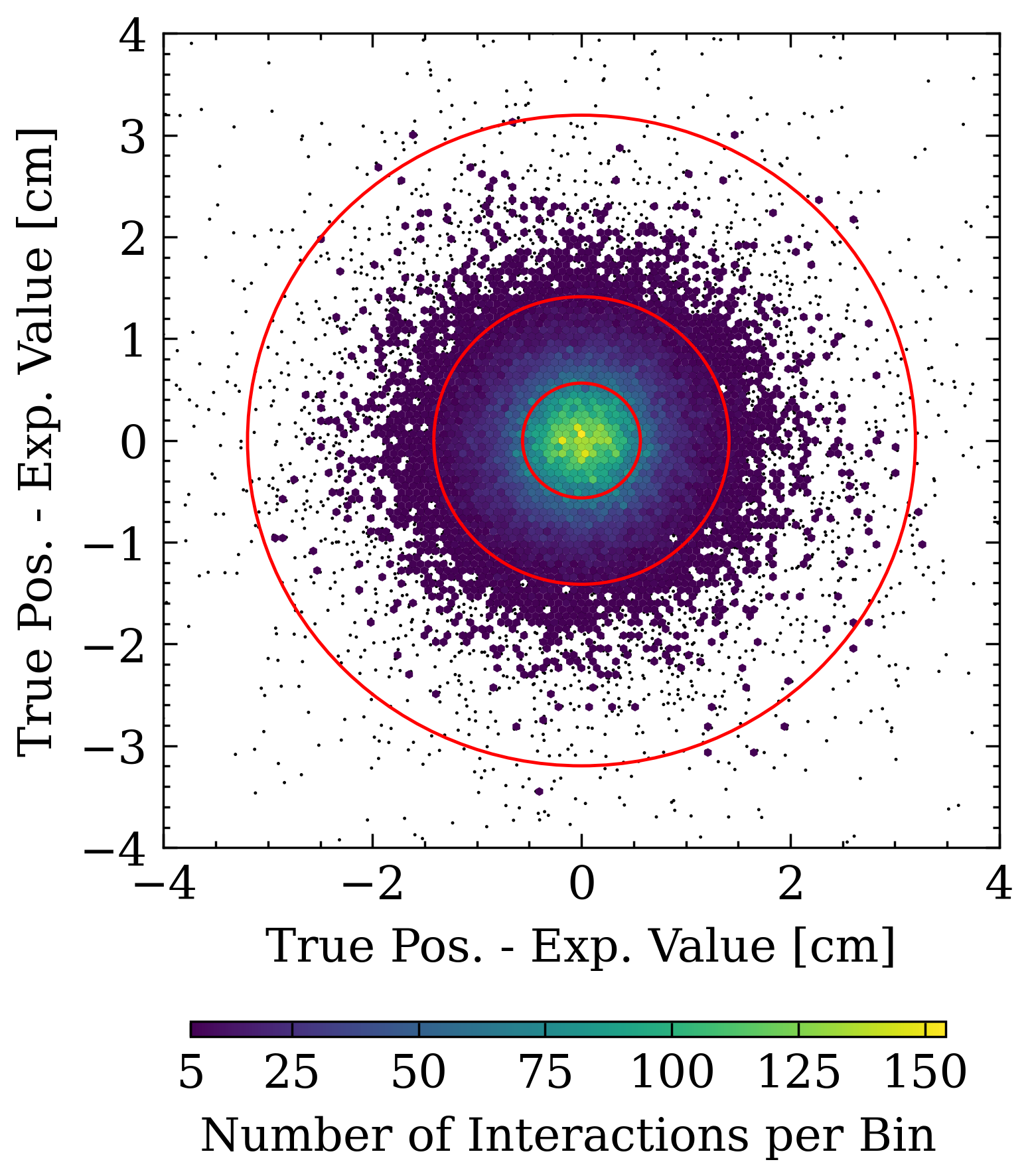}
\caption{Difference between the ground-truth position and expectation value of position of the interactions in the test set, shown in Cartesian coordinates as a two-dimensional histogram. 
The radii of the 1, 2, 3-$\sigma$ levels, which are shown as red circles, are listed in Table~\ref{tab:results-Diffxy}.
The RMS of ${\Delta x}$ and ${\Delta y}$ is 0.751~cm and 0.755~cm.
\label{fig:results-Diffxy}}
\end{figure}

Figure~\ref{fig:results-Diffxy} shows a two-dimensional histogram of the test set data with ${\Delta x}$ on the $x$-axis and ${\Delta y}$ on the $y$-axis.
In \cite{Liang2022} the authors report the performance of a Domain-informed Neural Network (DiNN) as the root mean square (RMS) of ${\Delta x}$ and ${\Delta y}$: 0.228~cm and 0.229~cm respectively. 
The RMS of ${\Delta x}$ and ${\Delta y}$ for this work is 0.751~cm and 0.755~cm.
This spread is dominated by the 1~cm$^2$ area of the discrete cells and could be reduced by increasing the number of cells.

\begin{table}[b]
\begin{tabular}{ |c|c|  }
 \hline
  Region & Radius \\
 \hline
  1-sigma (0.393) & 0.564~cm \\ 
  \hline
  2-sigma (0.865) & 1.413~cm \\  
  \hline
  3-sigma (0.989) & 3.196~cm \\ 
 \hline
\end{tabular}
\caption{\label{tab:results-Diffxy} 
Figure~\ref{fig:results-Diffxy} Confidence Regions.}
\end{table}

\subsection{Position Reconstruction at Detector Wall} \label{sec:results-detectorwall}

\begin{figure*}
\begin{center}
\begin{tabular}{cc}
\includegraphics[width=0.46\textwidth]{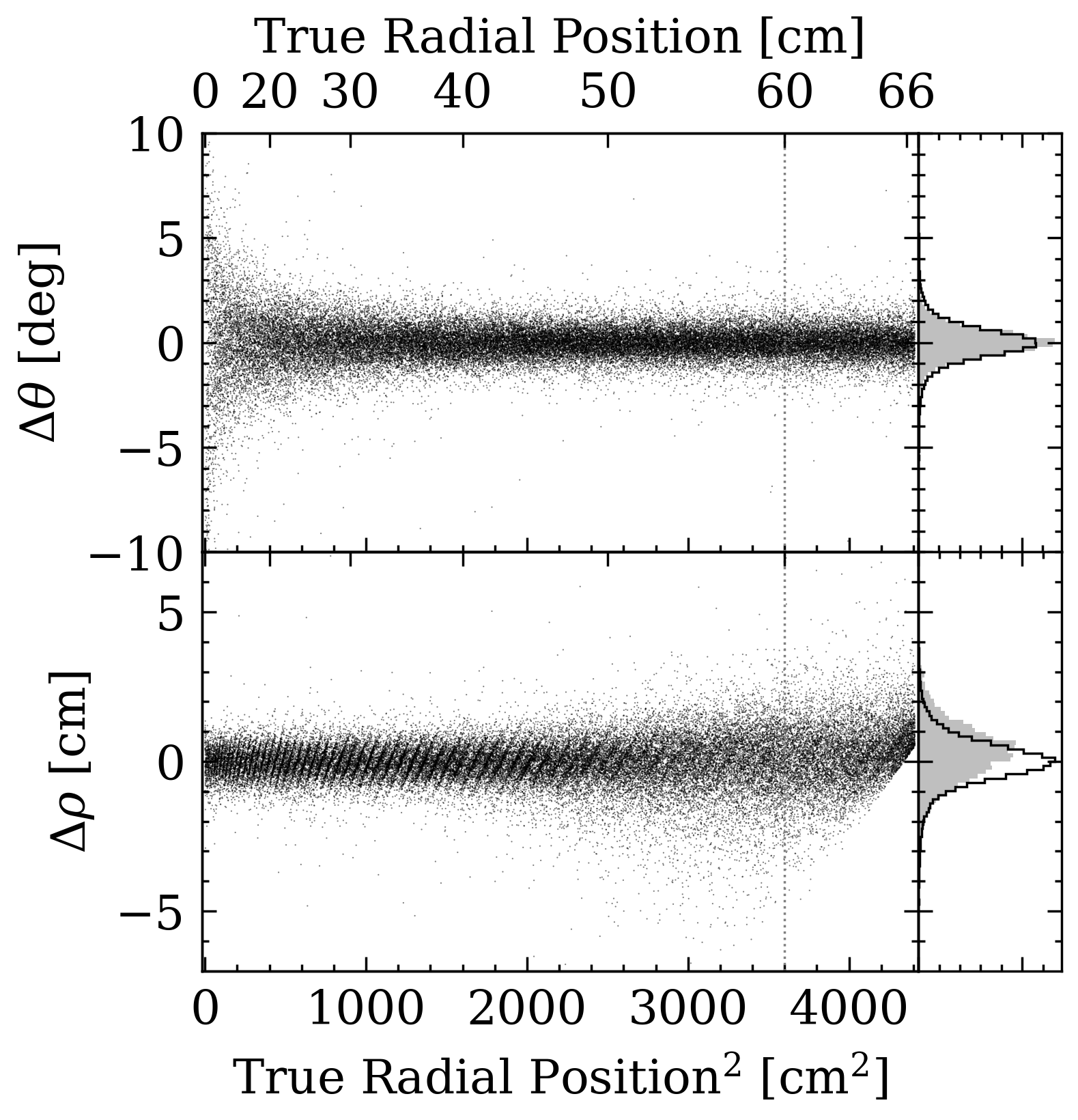} &
\includegraphics[width=0.46\textwidth]{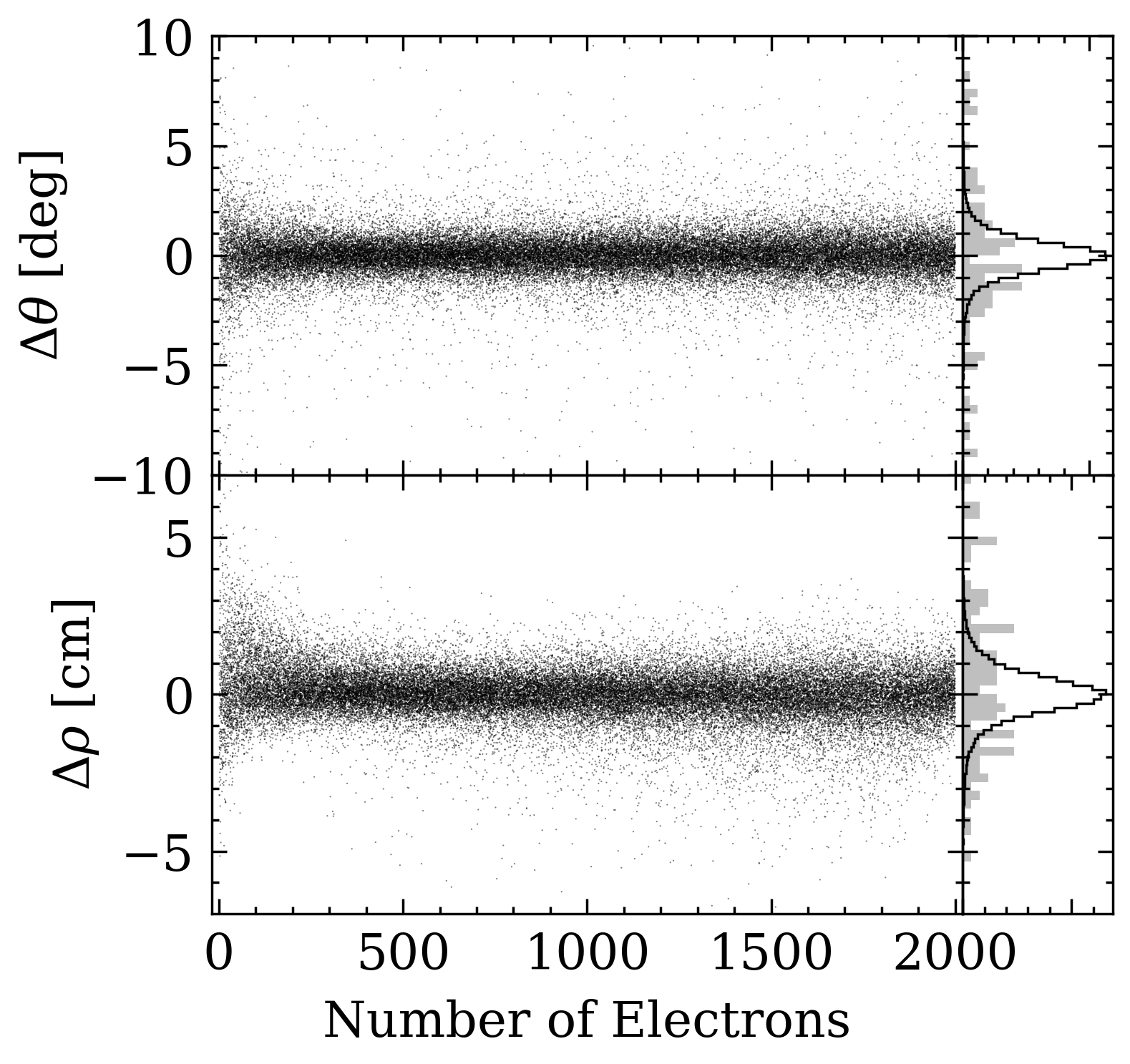}
\end{tabular}
\caption{Difference between the ground-truth position and expectation value position in polar coordinates, as a function of the ground-truth radial position (left) and ground-truth number of electrons (right). 
The diagonal lines in the bottom panel of the left figure are due to the discrete binning in position.
In the right panels of both figures, the black line projection histograms show the distribution for all data points.
The filled grey projection histograms show the distribution for true radial position $\geq 60$~cm in the left figure and for number of electrons $\leq 5$ in the right figure.
\label{fig:results-Diffrhophi}}
\end{center}
\end{figure*}

Current methods of position reconstruction perform less accurately at large radii, near the edge of the detector~\cite{Lux2018,XENON2019analysis}.
The fiducial volume, the region of interest within the detector, is defined by removing the region near the wall where the backgrounds are greatest.
For this reason, trends in the accuracy along the radial direction are of particular interest.
To compare the accuracy of the position reconstruction we calculate the difference between the ground-truth location and expectation value of position along the radial and angular axes,
\begin{equation}
\Delta \rho^i = \rho^i - {\langle \rho \rangle}^i
\ \mathrm{ and }\ 
\Delta \phi^i = \phi^i - {\langle \phi \rangle}^i.
\end{equation}
\noindent A larger spread of $\Delta \rho$ values near the wall than in the center would indicate the Bayesian network position reconstruction is performing less accurately near the wall of the detector.
In the top panel of Figure~\ref{fig:results-Diffrhophi}, we show $\Delta \phi$ as a function of the true radial position. 
Notice that the center of the $\Delta \phi$ distribution remains centered at zero for all values of true radial position.
In the bottom panel of Figure~\ref{fig:results-Diffrhophi}, we show that near the wall of the detector the mean of the $\Delta \rho$ distribution shifts upward $\sim$ 0.75~cm and the spread increases.
In the $\rho\ <\ $60~cm region, the interactions are reconstructed with an RMS error of 0.797~cm and at $\rho\ \geq\ $60~cm the RMS error is 1.108~cm.

\begin{figure*}
\begin{center}
\begin{tabular}{cc}
\includegraphics[width=0.46\textwidth]{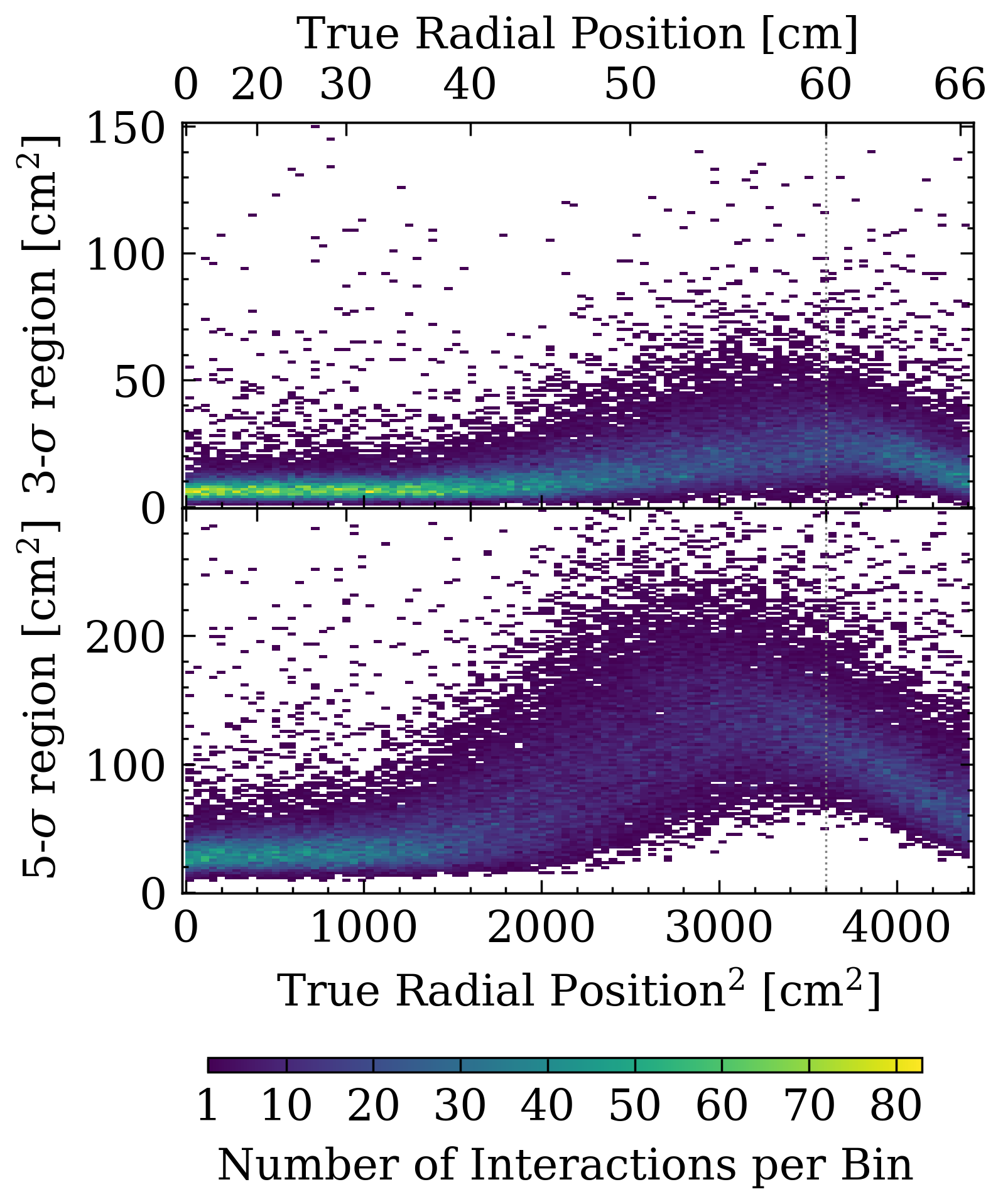} &
\includegraphics[width=0.46\textwidth]{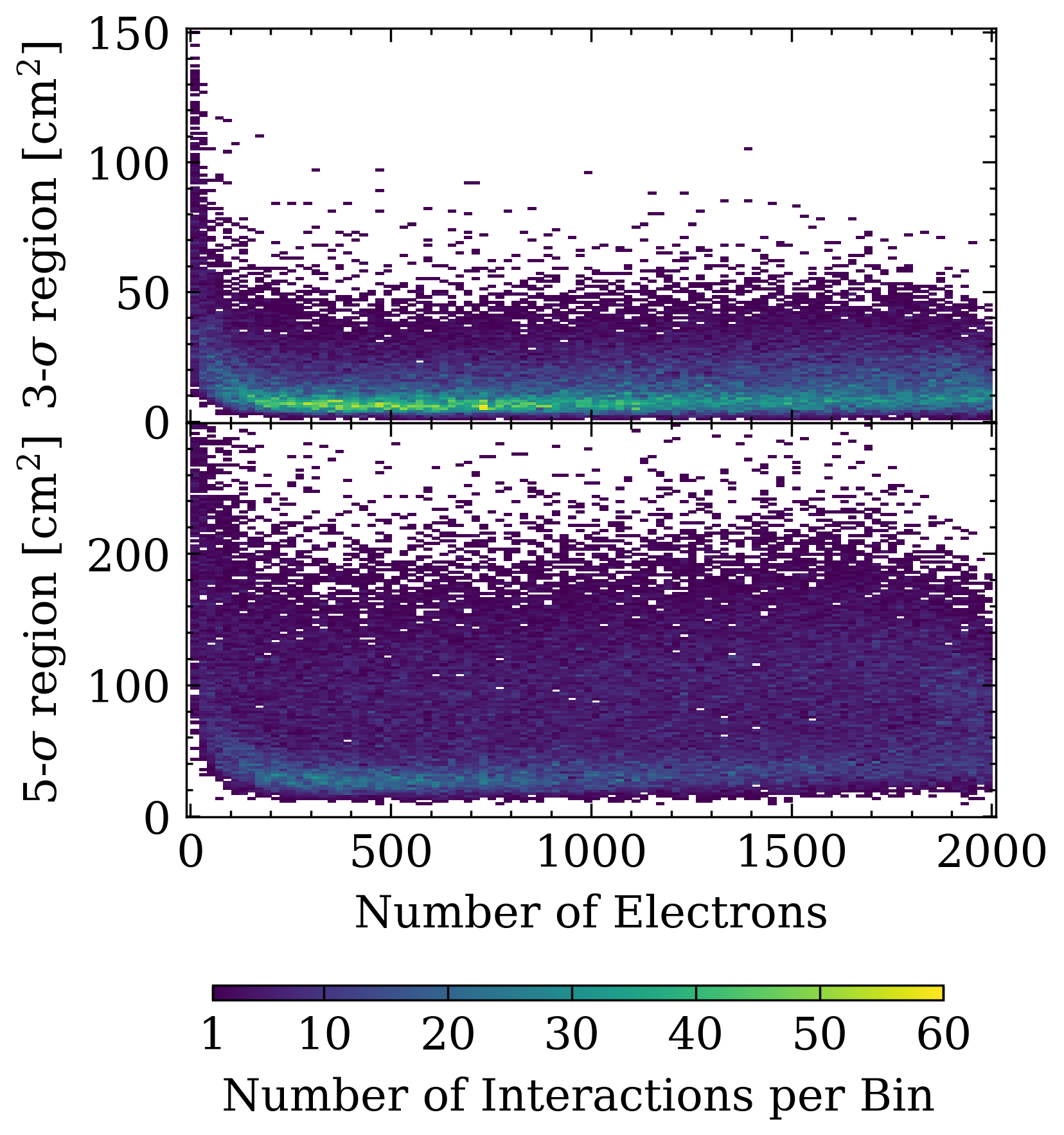}
\end{tabular}
\caption{Area of the confidence intervals, as a function of the ground-truth radial position (left) and ground-truth number of electrons (right). 
The areas of the 3-$\sigma$ confidence intervals (containing at least 98.8891\% of the probability) and the area of the 5-$\sigma$ confidence interval (containing at least 99.9996\% of the probability) are shown in the top and bottom panels of each figure.
The data is shown as a 2D histogram where the color indicates the number of interactions in each bin.
\label{fig:results-confidencesize}}
\end{center}
\end{figure*}

Additionally, larger areas for the 3-$\sigma$ confidence near the wall than in the center would indicate that the reconstruction is less certain near the wall of the detector.
We show, in Figure~\ref{fig:results-confidencesize}, the area of the 3-$\sigma$ and 5-$\sigma$ confidence regions as a function of the true radial position.
The position reconstruction confidence region is largest at true radial positions between 50~cm and 60~cm, then decreases for $\rho\ \geq\ $60~cm.
Finally, the RMS of ${\Delta x}$ and ${\Delta y}$ (shown in Figure~\ref{fig:results-Diffxy}) is 0.692~cm and 0.697~cm for $\rho\ <\ $60~cm and 0.973 and 0.974~cm for $\rho\ \geq\ $60~cm.

\subsection{Position Reconstruction of Interactions with Few Electron} \label{sec:results-fewelectrons}

Current methods of position reconstruction perform less accurately on low energy interactions.
Similarly to the metric discussed in Section~\ref{sec:results-detectorwall}, we consider $\Delta \rho$ and $\Delta \phi$, as well as the area of the 3-$\sigma$ and 5-$\sigma$ confidence regions, but as a function of the ground-truth number of electrons.
The right panel of Figure~\ref{fig:results-Diffrhophi} shows there is a greater spread in both $\Delta \rho$ and $\Delta \phi$ for low energy interactions.
In the $E \leq 5$ region the interactions are reconstructed with an RMS error of 2.809~cm and for $E > 5$ the RMS error is 0.854~cm.
Similarly, the right panel of Figure~\ref{fig:results-confidencesize} shows that both the 3-$\sigma$ and 5-$\sigma$ confidence regions have a much larger area for $E \leq 5$.
In the $E \leq 5$ region, the median 3-$\sigma$ confidence region area is 12~cm$^2$, and for $E > 5$ the median 3-$\sigma$ confidence region area is 171~cm$^2$.
This is due to the hit patterns of the $E \leq 5$ interactions being more impacted by the Poisson fluctuations than the $E > 5$ interactions.

\section{Conclusions} \label{sec:conclusions}

In this work, we have developed for the first time a method for position reconstruction using a Bayesian network and demonstrated its utility using a dual-phase TPC as a proof-of-concept.
The Bayesian network used a graph structure with nodes representing the interaction position, the number of electrons that produced the second scintillation light, and the observed hit pattern.
This method is well suited to the challenge of position reconstruction and provides an interpretable model of the problem that represents the uncertainties within the processes as probability distributions.

The reconstruction precision achieved is comparable to the state-of-the-art methods \cite{Lux2018, XENON2019analysis, Liang2022, PandaX2021}.
A test set of 50,000 particle interactions were simulated based on the XENONnT detector geometry.
The interaction positions were generated from a uniform random distribution over the entire detector area and the number of electrons extracted into the gas were generated from a uniform random distribution from 1 to 2000.
For the inner part of the detector ($<$60~cm), an RMS of 0.69~cm was achieved ($\sim$0.09 of the sensor spacing), whereas near the wall of the detector ($\geq$60~cm), an RMS of $\sim$0.98~cm was achieved ($\sim$0.12 of the sensor spacing).
The RMS is dominated by the 1~cm$^2$ area of the discrete cells and could be reduced by increasing the number of cells.

Often the RMS of the sample, which is a measure of how far on average the error is from zero, is reported as the position uncertainty.
Our method provides not only an RMS competitive with current methods, but more importantly, per interaction uncertainty, which is not possible with other reconstruction methods.
The median of the per interaction 3-$\sigma$ confidence region is 11~cm$^2$ for interactions in the inner part of the detector ($<$60~cm) and 21~cm$^2$ near the wall of the detector ($\geq$60~cm).
Moreover, the full posterior probability distributions over position can be propagated through a probabilistic analysis.
For example, the probability distribution could be used to define a probabilistic fiducial volume, which could further reduce background by removing mismeasurements.

A feature of this method is that the discretization of the detector area naturally constrains the position reconstruction to be within the physical volume of the detector, in contrast to neural networks which often produced nonphysical results or require a customized layer to constrain the output to be within the physical volume~\cite{Liang2022}.

In the future, we aim to incorporate reconstruction uncertainty into the search for neutrinoless double beta decay while increasing the robustness and interpretability of the dark matter searches in our field.
This method can be extended to energy reconstruction and three dimensional position reconstruction, as well as for signal classification, providing per interaction uncertainties.
These applications will require advances to our methodology, including more complex graph structures or learning the graph structure directly from the data.

\begin{acknowledgments}

This paper is dedicated to our friend, the late Professor Hagit Shatkay.
The culmination of her original ideas, this work is based on her discussions with CT and WB, which led to formation of this collaboration, and was then led by CP.

This work is supported by the National Science Foundation through awards 1940074, 1940209, and 1940080.

The authors would like to acknowledge the following open source projects which were used in this work: {\em numpy} \cite{Harris2020}, {\em scipy} \cite{SciPy2020}, and {\em matplotlib} \cite{Hunter2007}.

Finally, the authors would like to thank Daniel Wenz, Sophia Farrell, and Ivy Li for their excellent feedback which greatly improved this manuscript.

\end{acknowledgments}

\bibliography{references.bib}

\appendix

\onecolumngrid   
\section{Example Events} \label{sec:appendix-examples}

\begin{figure*}
\includegraphics[height=0.26\textheight]{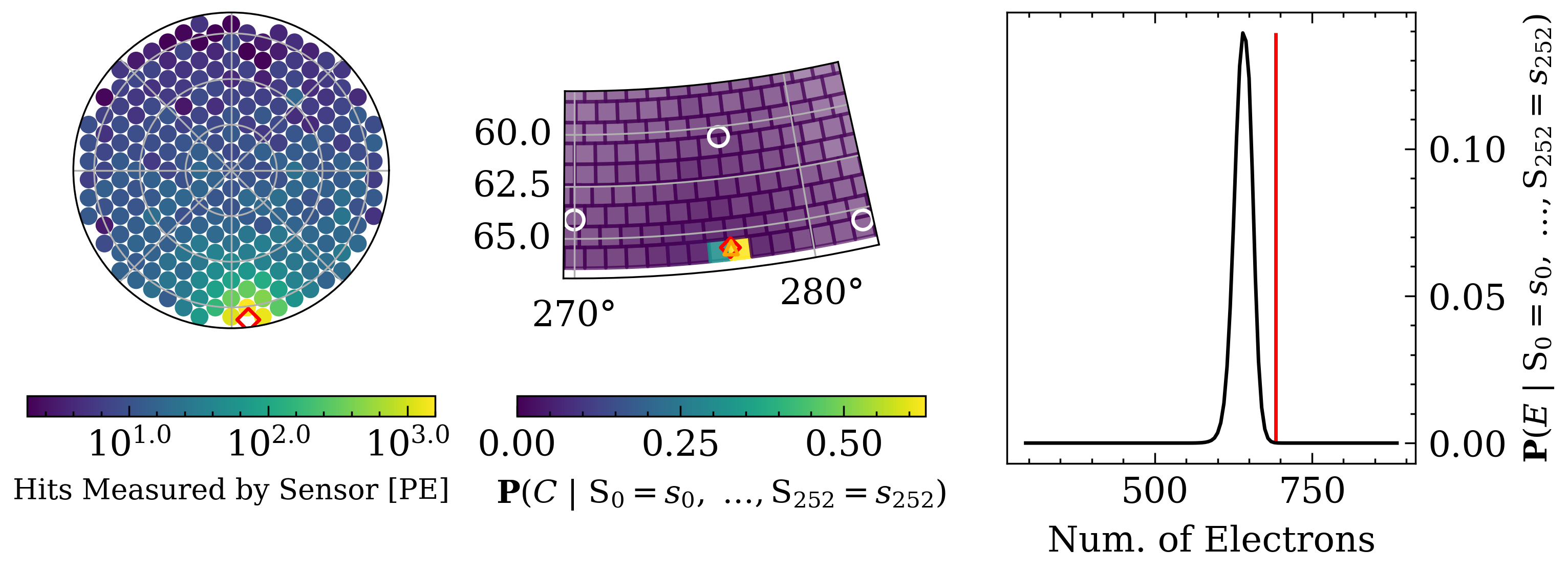}
\caption{
As described in Figure~\ref{fig:methods-examplehitpatternposterior}.
The difference between the true position and the expectation value of position is 0.03 cm.
The 99.0372\% ($\sim$3-$\sigma$) and 99.9996\% ($\sim$5-$\sigma$) confidence intervals are 8 cm$^2$ and 45 cm$^2$.
The ground truth number of electrons is 692.
\label{fig:appendix-event0}}
\end{figure*}

\begin{figure*}
\includegraphics[height=0.26\textheight]{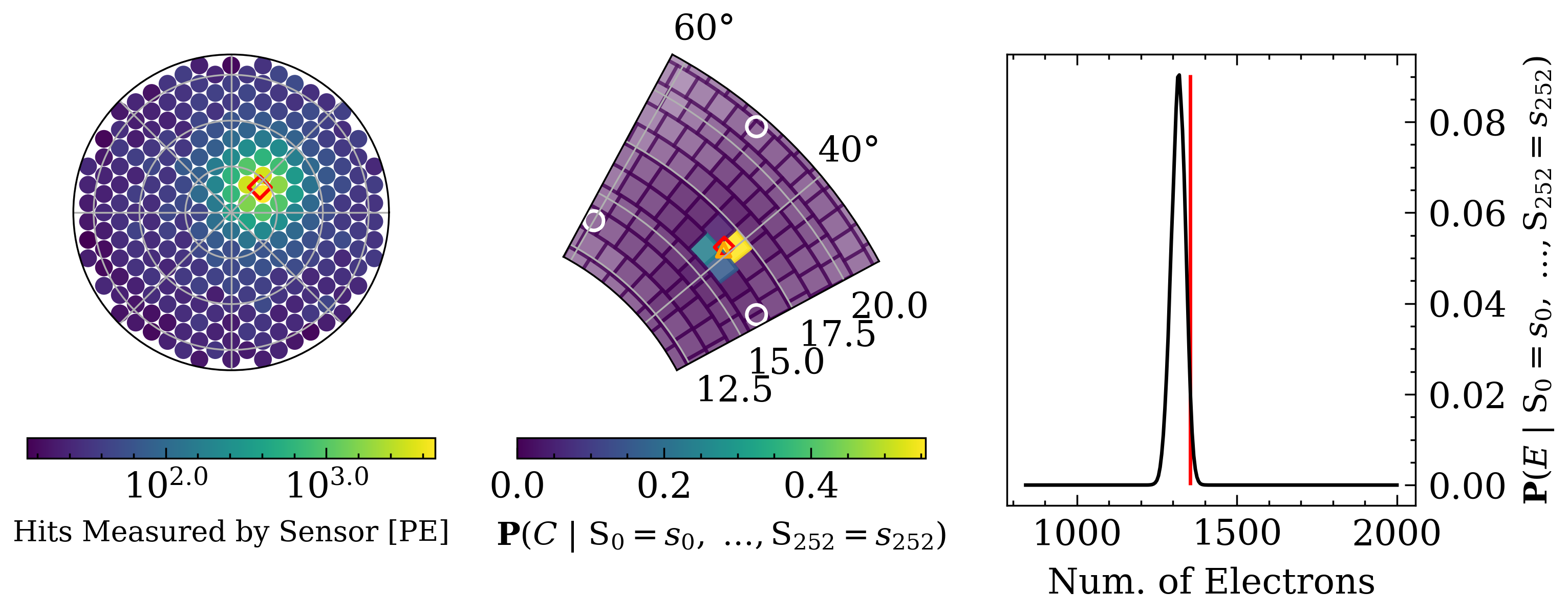}
\caption{
As described in Figure~\ref{fig:methods-examplehitpatternposterior}.
The difference between the true position and the expectation value of position is 0.12 cm.
The 99.5080\% ($\sim$3-$\sigma$) and 99.9997\% ($\sim$5-$\sigma$) confidence intervals are 5 cm$^2$ and 23 cm$^2$.
The ground truth number of electrons is 1354.
\label{fig:appendix-event1}}
\end{figure*}

\begin{figure*}
\includegraphics[height=0.26\textheight]{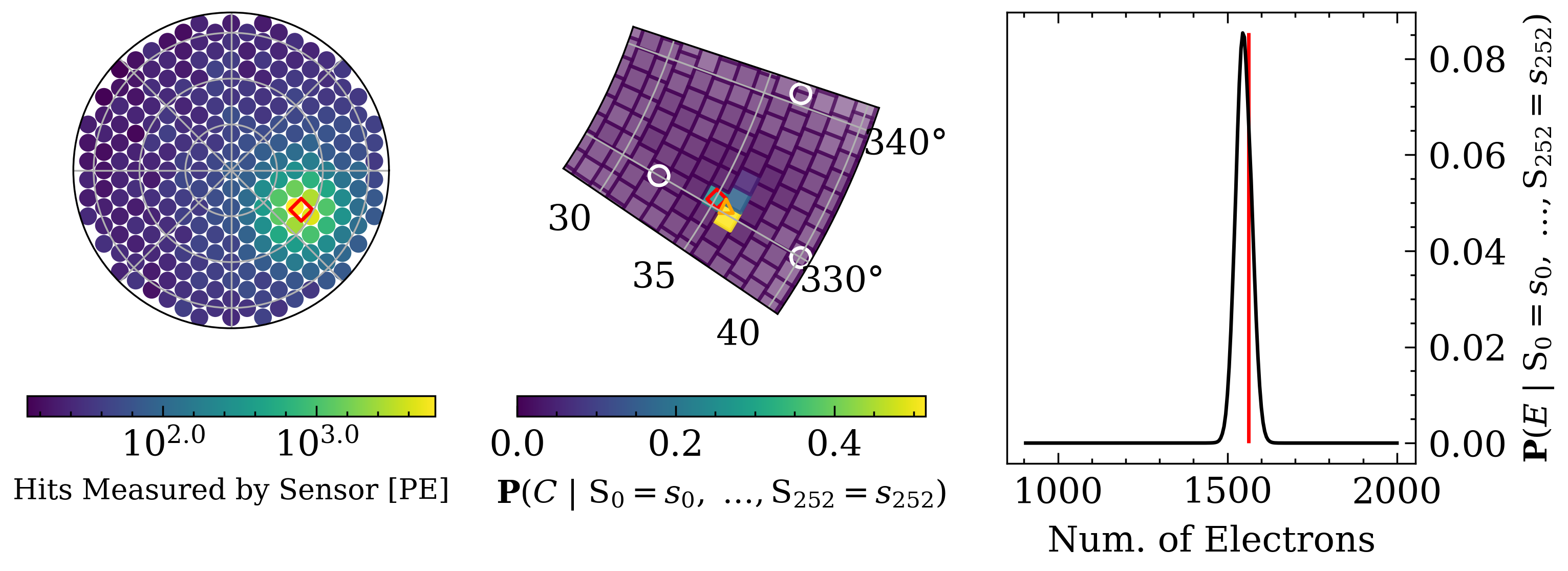}
\caption{
As described in Figure~\ref{fig:methods-examplehitpatternposterior}.
The difference between the true position and the expectation value of position is 0.58 cm.
The 99.2380\% ($\sim$3-$\sigma$) and 99.9996\% ($\sim$5-$\sigma$) confidence intervals have areas of 6 cm$^2$ and 39 cm$^2$.
The ground truth number of electrons is 1562.
\label{fig:appendix-event2}}
\end{figure*}

\begin{figure*}
\includegraphics[height=0.26\textheight]{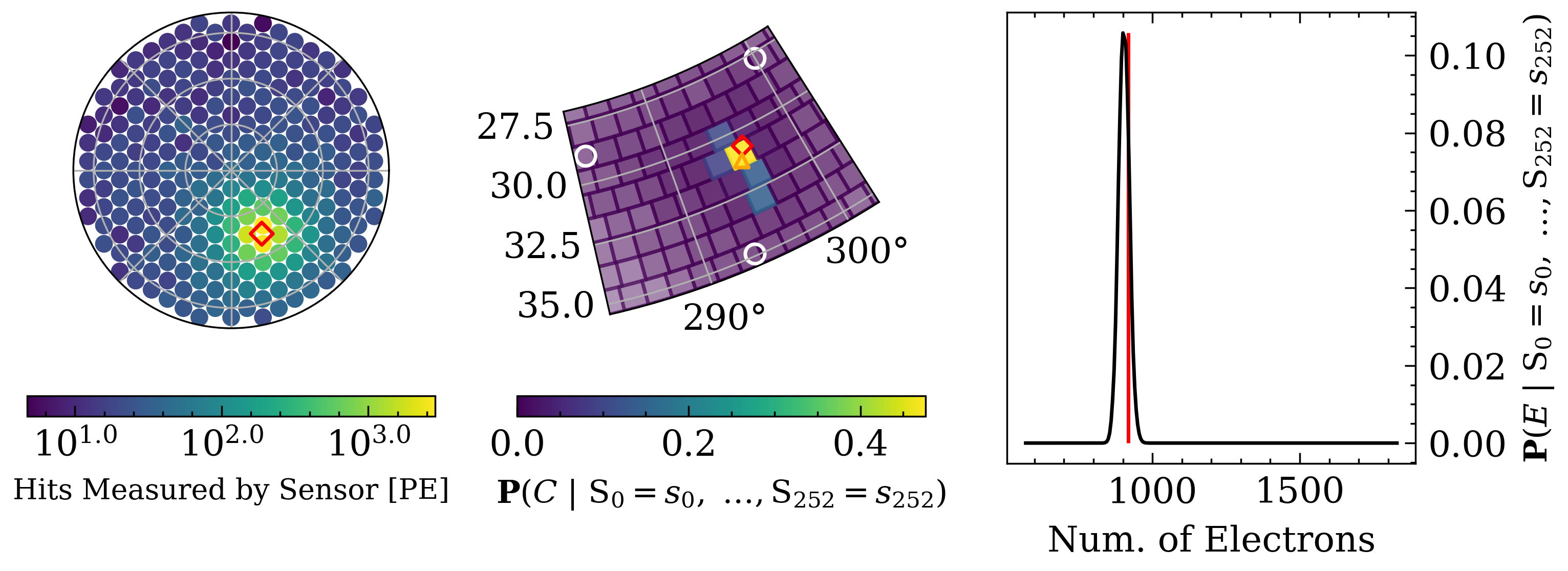}
\caption{
As described in Figure~\ref{fig:methods-examplehitpatternposterior}.
The difference between the true position and the expectation value of position is 0.61 cm.
The 99.1381\% ($\sim$3-$\sigma$) and 99.9997\% ($\sim$5-$\sigma$) confidence intervals have areas of 9 cm$^2$ and 36 cm$^2$.
The ground truth number of electrons is 919.
\label{fig:appendix-event3}}
\end{figure*}

\begin{figure*}
\includegraphics[height=0.26\textheight]{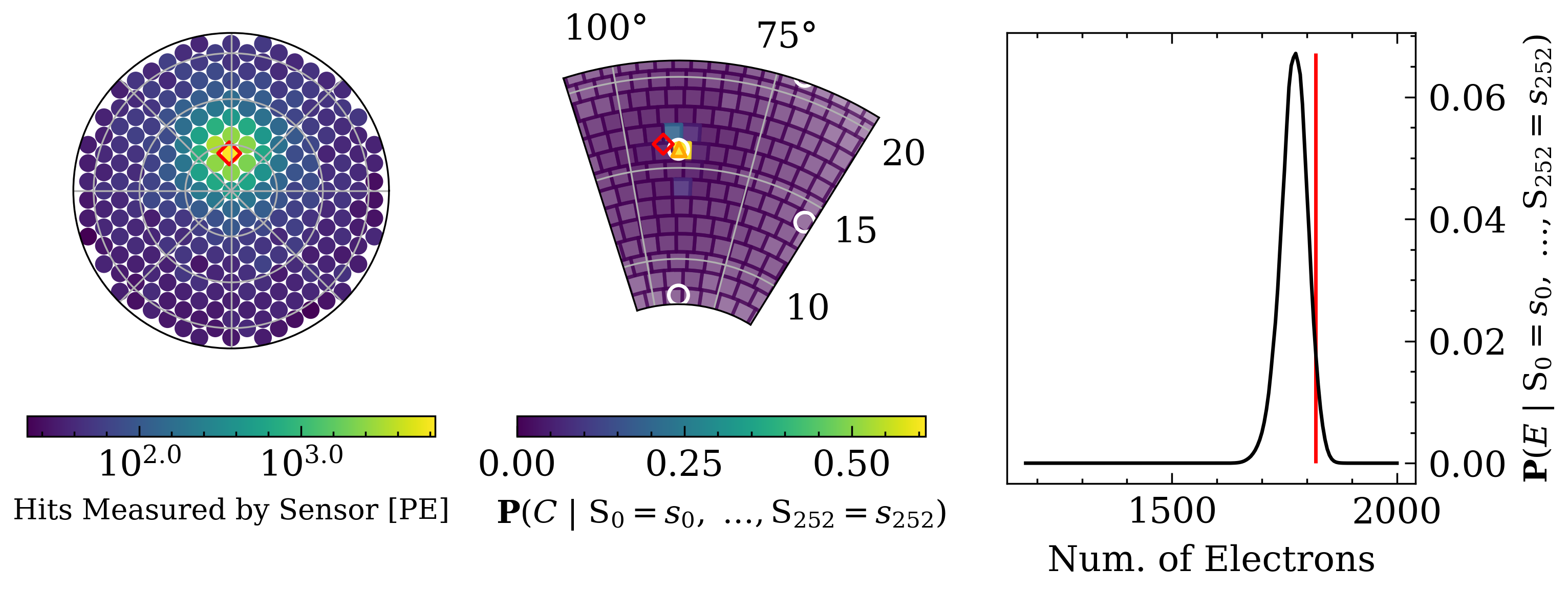}
\caption{
As described in Figure~\ref{fig:methods-examplehitpatternposterior}.
The difference between the true position and the expectation value of position is 0.91 cm.
The 99.0957\% ($\sim$3-$\sigma$) and 99.9997\% ($\sim$5-$\sigma$) confidence intervals have areas of 9 cm$^2$ and 51 cm$^2$.
The ground truth number of electrons is 1820.
\label{fig:appendix-event4}}
\end{figure*}

\begin{figure*}
\includegraphics[height=0.26\textheight]{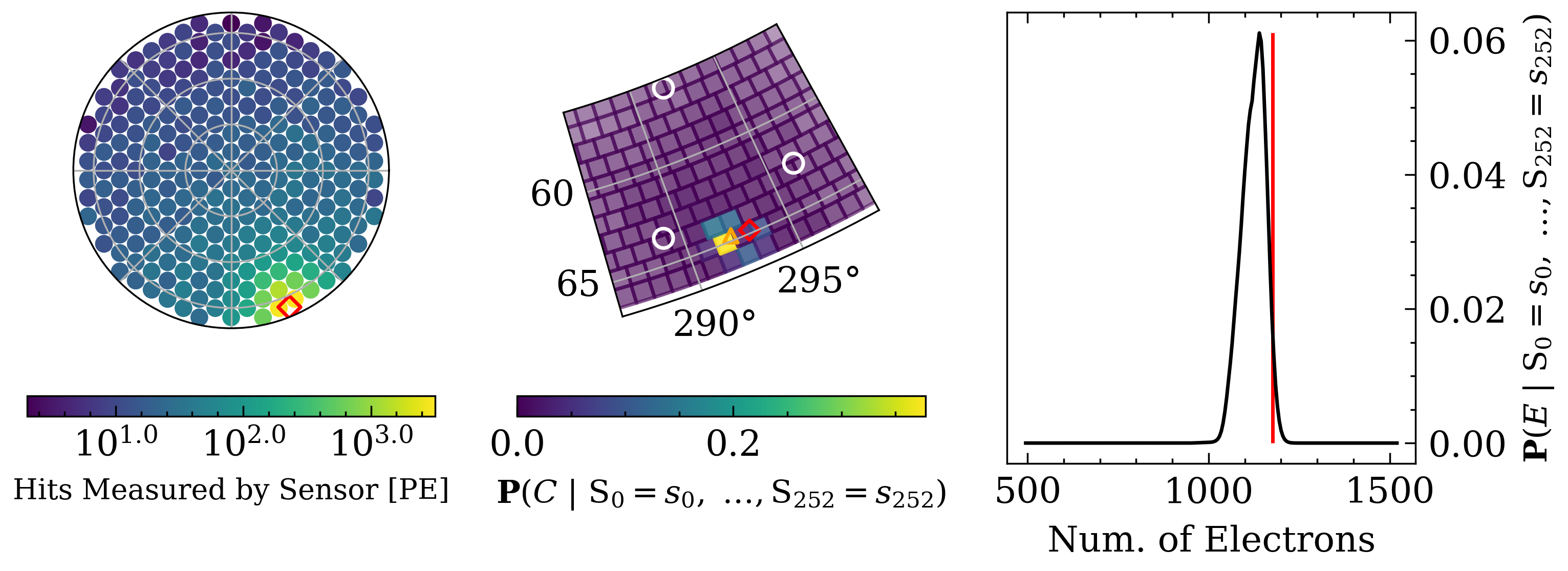}
\caption{
As described in Figure~\ref{fig:methods-examplehitpatternposterior}.
The difference between the true position and the expectation value of position is 1.04 cm.
The 98.9500\% ($\sim$3-$\sigma$) and 99.9997\% ($\sim$5-$\sigma$) confidence intervals have areas of 19 cm$^2$ and 92 cm$^2$.
The ground truth number of electrons is 1177.
\label{fig:appendix-event5}}
\end{figure*}

\begin{figure*}[p]
\includegraphics[height=0.26\textheight]{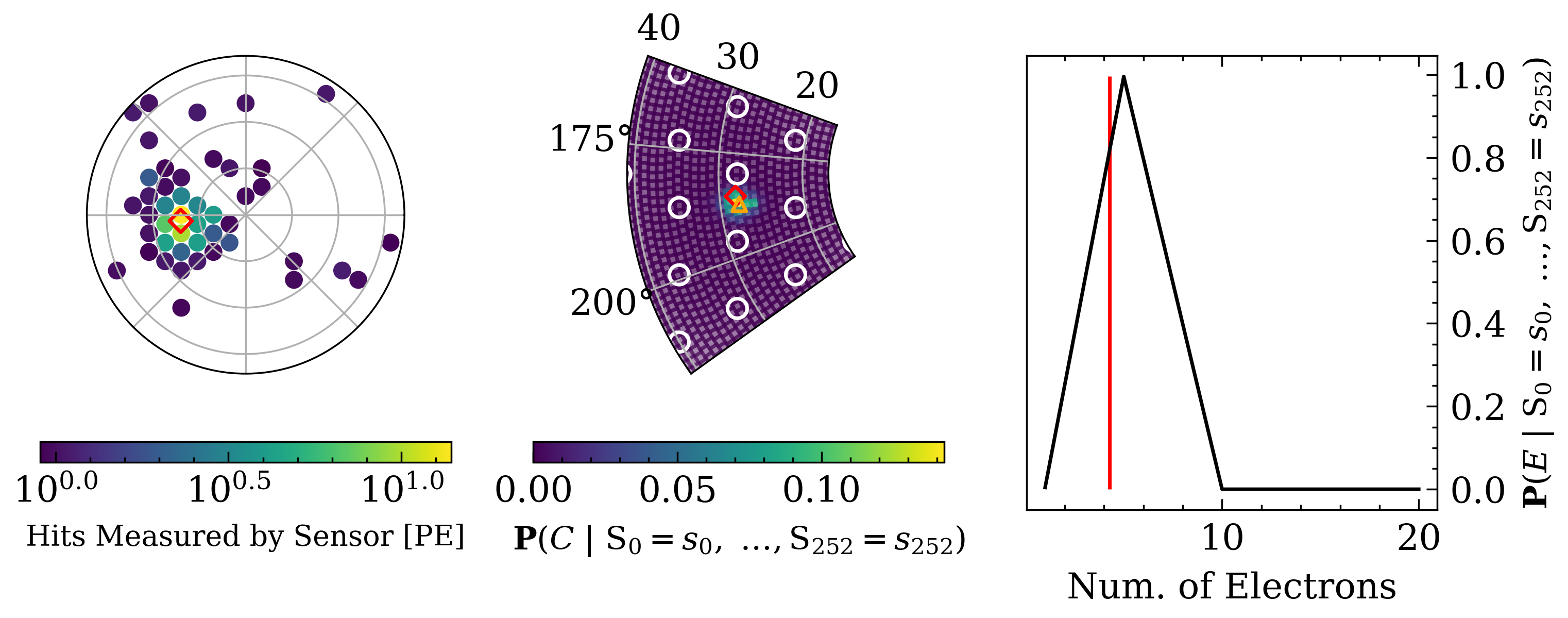}
\caption{
As described in Figure~\ref{fig:methods-examplehitpatternposterior}.
The difference between the true position and the expectation value of position is 1.06 cm.
The 98.8891\% ($\sim$3-$\sigma$) and 99.9996\% ($\sim$5-$\sigma$) confidence intervals have areas of 38 cm$^2$ and 471 cm$^2$.
The ground truth number of electrons is 4.
\label{fig:appendix-event6}}
\end{figure*}

\begin{figure*}
\includegraphics[height=0.26\textheight]{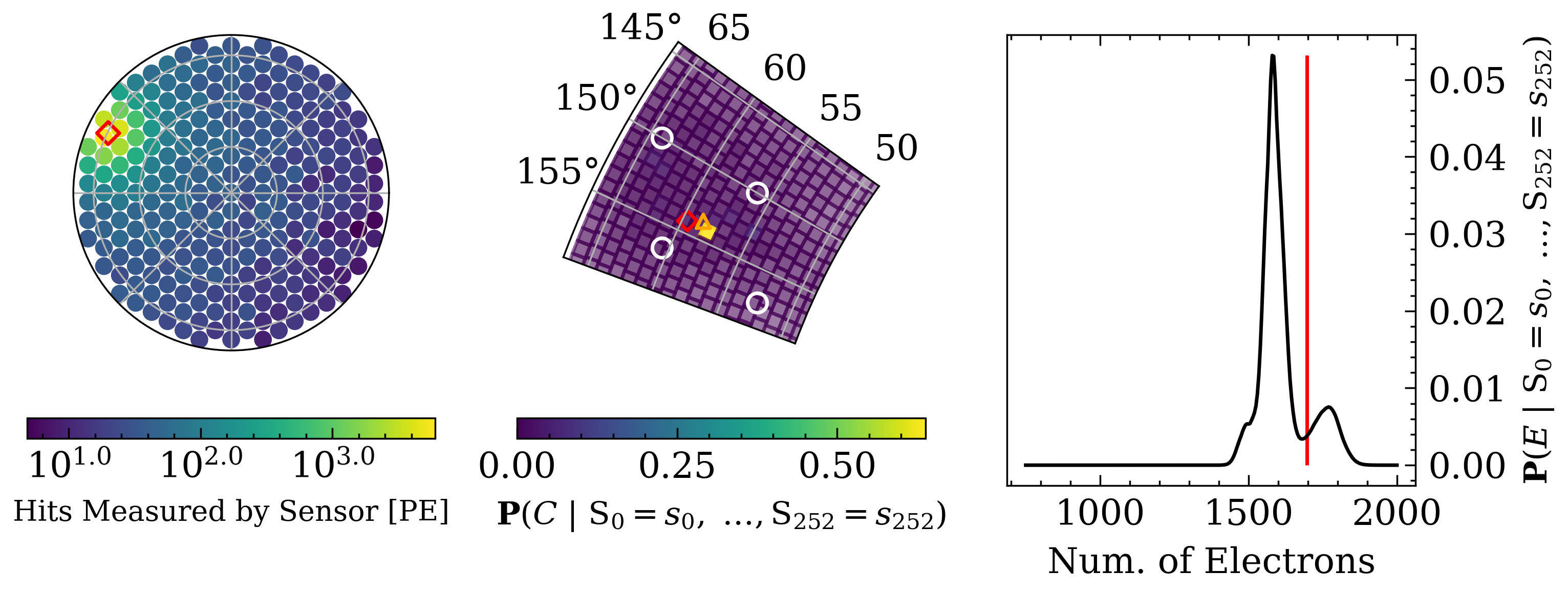}
\caption{
As described in Figure~\ref{fig:methods-examplehitpatternposterior}.
The difference between the true position and the expectation value of position is 1.16 cm.
The 98.9188\% ($\sim$3-$\sigma$) and 99.9996\% ($\sim$5-$\sigma$) confidence intervals have areas of 43 cm$^2$ and 154 cm$^2$.
The ground truth number of electrons is 1697.
\label{fig:appendix-event7}}
\end{figure*}

\begin{figure*}
\includegraphics[height=0.26\textheight]{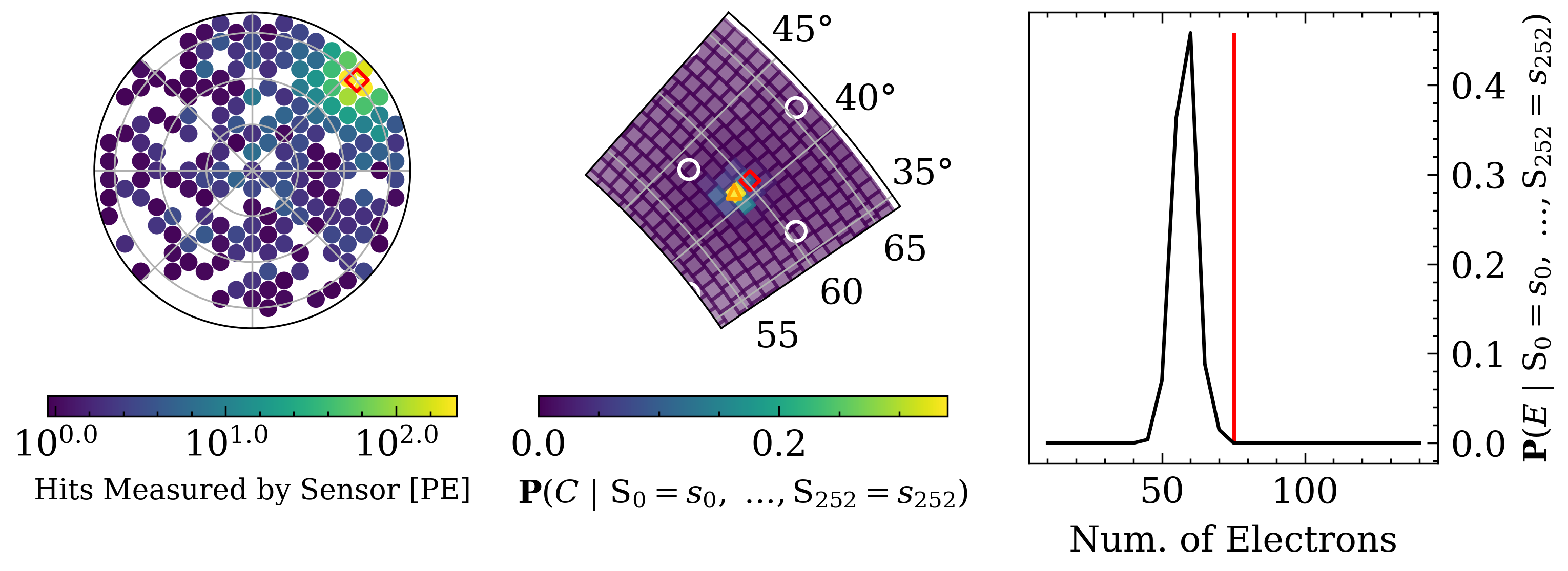}
\caption{
As described in Figure~\ref{fig:methods-examplehitpatternposterior}.
The difference between the true position and the expectation value of position is 1.26 cm.
The 99.0147\% ($\sim$3-$\sigma$) and 99.9996\% ($\sim$5-$\sigma$) confidence intervals have areas of 21 cm$^2$ and 126 cm$^2$.
The ground truth number of electrons is 75.
\label{fig:appendix-event8}}
\end{figure*}

\begin{figure*}
\includegraphics[height=0.26\textheight]{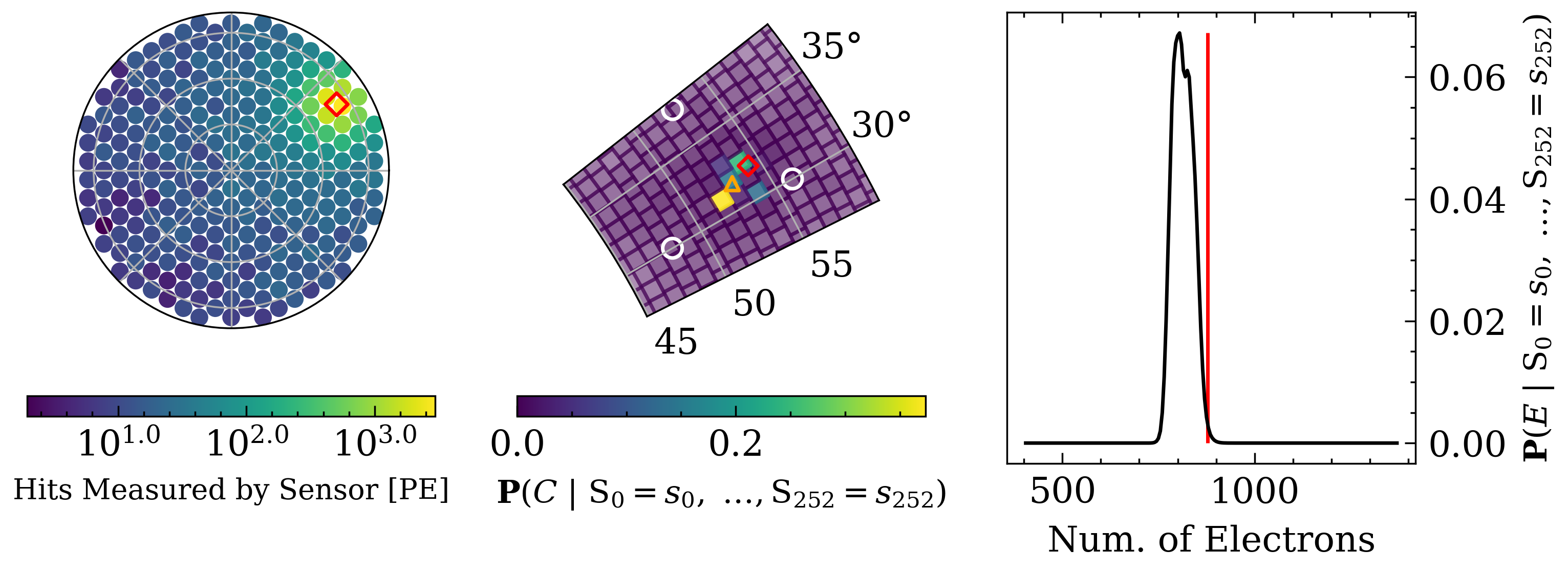}
\caption{
As described in Figure~\ref{fig:methods-examplehitpatternposterior}.
The difference between the true position and the expectation value of position is 1.38 cm.
The 98.9041\% ($\sim$3-$\sigma$) and 99.9996\% ($\sim$5-$\sigma$) confidence intervals have areas of 11 cm$^2$ and 75 cm$^2$.
The ground truth number of electrons is 878.
\label{fig:appendix-event9}}
\end{figure*}

\begin{figure*}
\includegraphics[height=0.26\textheight]{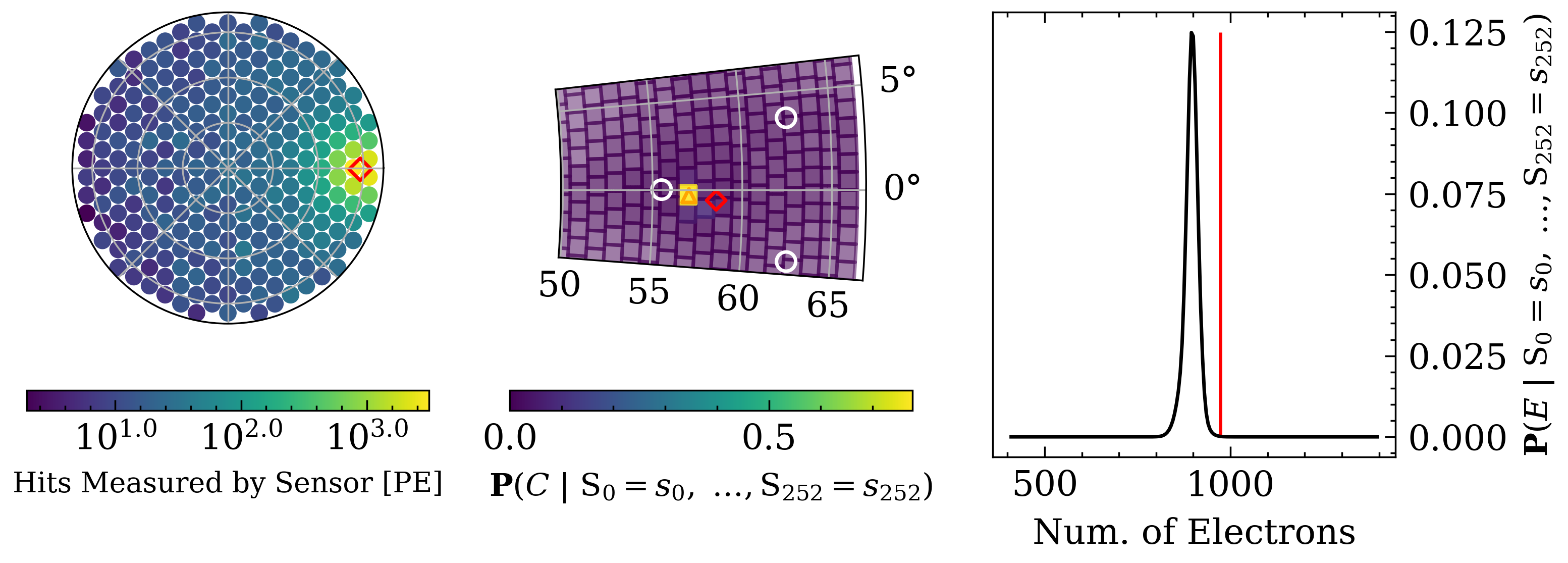}
\caption{
As described in Figure~\ref{fig:methods-examplehitpatternposterior}.
The difference between the true position and the expectation value of position is 1.54 cm.
The 99.0891\% ($\sim$3-$\sigma$) and 99.9996\% ($\sim$5-$\sigma$) confidence intervals have areas of 11 cm$^2$ and 109 cm$^2$.
The ground truth number of electrons is 972.
\label{fig:appendix-event10}}
\end{figure*}

\begin{figure*}
\includegraphics[height=0.26\textheight]{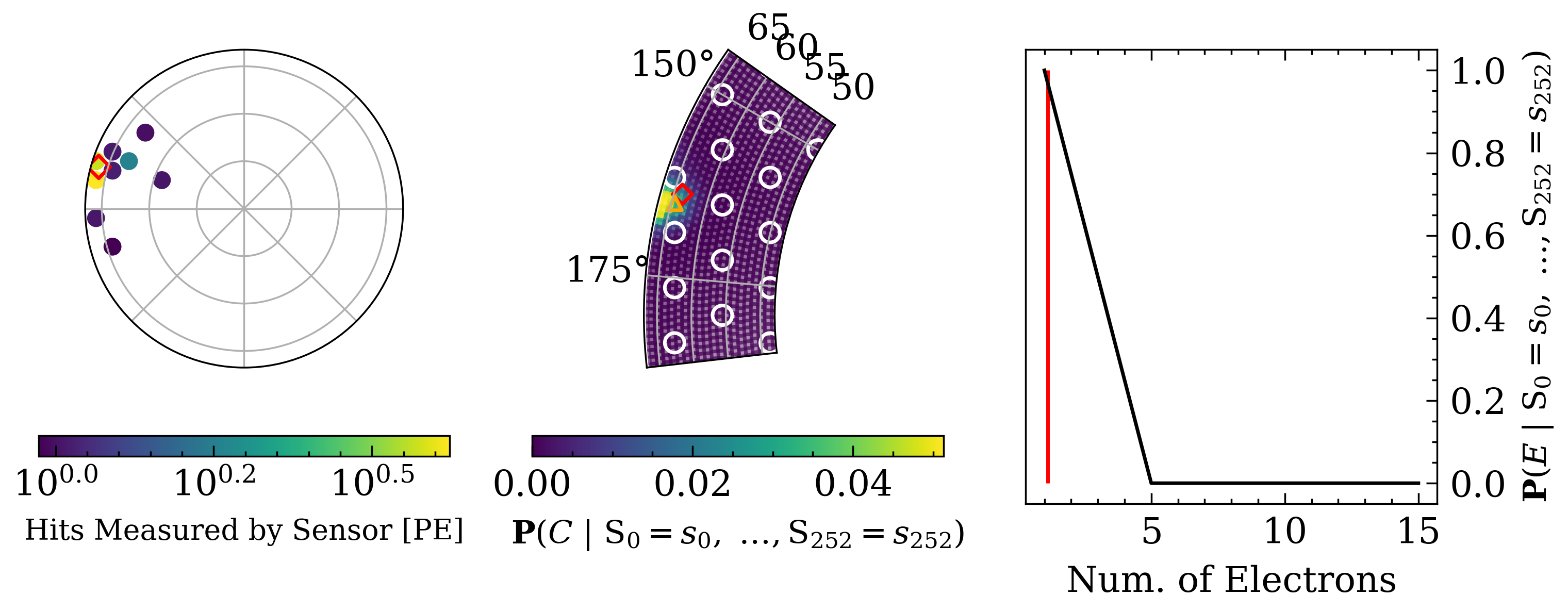}
\caption{
As described in Figure~\ref{fig:methods-examplehitpatternposterior}.
The difference between the true position and the expectation value of position is 1.71 cm.
The 98.9307\% ($\sim$3-$\sigma$) and 99.9996\% ($\sim$5-$\sigma$) confidence intervals have areas of 93 cm$^2$ and 435 cm$^2$.
The ground truth number of electrons is 1.
\label{fig:appendix-event11}}
\end{figure*}

\end{document}